\documentclass[floatfix,superscriptaddress,longbibliography,nofootinbib]{revtex4-2}
\RequirePackage[sort&compress]{natbib}
\usepackage{color}
\usepackage{float}
\usepackage[table]{xcolor}
\usepackage{amssymb,amsfonts,amsmath}
\usepackage{bm}
\usepackage[pdftex]{graphicx}
\usepackage{xcolor}
\usepackage{comment}
\usepackage{multirow}
\usepackage{booktabs}

\usepackage[english]{babel}
\usepackage{amsthm}

\theoremstyle{definition}
\newtheorem{definition}{Definition}[section]

\usepackage[utf8]{inputenc} 
\usepackage[T1]{fontenc}    
\usepackage{url}            
\usepackage{booktabs}       
\usepackage{amsfonts}       
\usepackage{nicefrac}       
\usepackage{microtype}      
\usepackage{xcolor}         
\usepackage{graphicx}
\usepackage{amsmath}
\usepackage{natbib}
\usepackage{longtable}
\usepackage{listings}
\usepackage{adjustbox}
\usepackage{array}
\usepackage{ragged2e} 
\usepackage{placeins}  
\usepackage{array}

\usepackage{dcolumn}
\usepackage{bm}

\usepackage{xcolor}

\usepackage{xcolor}
\definecolor{RoyalBlue}{HTML}{4169e1}
\definecolor{ForestGreen}{HTML}{228b22}

\usepackage{setspace}
\usepackage{comment}

\begin{document}

\title{Large Language Models are Zero-Shot Next Location Predictors}

\author{Ciro Beneduce}
\affiliation{Fondazione Bruno Kessler, Via Sommarive 18, 38123 Povo (TN), Italy}

\author{Bruno Lepri}
\affiliation{Fondazione Bruno Kessler, Via Sommarive 18, 38123 Povo (TN), Italy}

\author{Massimiliano Luca}
\email[Corresponding author:~]{mluca@fbk.eu}%
\affiliation{Fondazione Bruno Kessler, Via Sommarive 18, 38123 Povo (TN), Italy}

\date{\today}

\begin{abstract}
Predicting the locations an individual will visit in the future is crucial for solving many societal issues like disease diffusion and reduction of pollution. However, next-location predictors require a significant amount of individual-level information that may be scarce or unavailable in some scenarios (e.g., cold-start). Large Language Models (LLMs) have shown good generalization and reasoning capabilities and are rich in geographical knowledge, allowing us to believe that these models can act as zero-shot next-location predictors. We tested more than 15 LLMs on three real-world mobility datasets and we found that LLMs can obtain accuracies up to 36.2\%, a significant relative improvement of almost 640\% when compared to other models specifically designed for human mobility. We also test for data contamination and explored the possibility of using LLMs as text-based explainers for next-location prediction, showing that, regardless of the model size, LLMs can explain their decision. 
\end{abstract}

\maketitle


\section*{Introduction}
Next-location prediction (NL) consists of predicting an individual's future whereabouts based on their historical visits to locations. Predicting future movements of individuals is crucial for addressing various societal challenges, including traffic management and optimization, control of disease diffusion, design of more sustainable urban spaces, disaster response management and others \citep{barbosa2018human,  yabe2023metropolitan, lee2015relating, bohm2022gross, akhtar2021review, wang2012understanding, wesolowski2012quantifying, xiong2020mobile, kraemer2020effect, yabe2023behavioral, moro2021mobility, lu2012predictability, deville2014dynamic, gonzalez2008understanding, bosetti2020heterogeneity, klamser2023}. 

NL has been faced using pattern-based \citep{trasarti2017myway,comito2017you}, Markov-based \citep{gambs2012next, bontorin2024mixing}, and Deep Learning (DL) based approaches \citep{feng2018deepmove,luca2020survey, liu2016predicting,sun2020go, luo2021stan, xue2021mobtcast}. Following the evolution of DL techniques, researchers shifted from solutions based on recurrent networks \citep{liu2016predicting,feng2018deepmove} to more sophisticated transformer-like architectures \citep{xue2021mobtcast}. These models can effectively predict the next location an individual will visit in most circumstances. However, there are scenarios in which even sophisticated DL models fail such predictions. Examples are out-of-routine mobility \citep{luca2023trajectory} and data-scarce geographical areas where transferable models or zero-shot learners are needed \citep{luca2020survey}. Thus, researchers have designed specific datasets for cross-city mobility studies \citep{yabe2023metropolitan} and for evaluating geographically transferable algorithms \citep{simini2021deep, mcarthur2011spatial}. However, the most prominent results have been obtained for collective mobility models as they depend on population density and distances between locations \citep{simini2021deep, mcarthur2011spatial};. At the same time, geographic transferability and zero-shot learning are still an open issue when it comes to individual behaviours \citep{luca2020survey}.

Recent progress in DL has led to the proliferation of Large Language Models (LLMs), which are now used in various domains, including time-series, finance, healthcare, and mobility forecasting \citep{jin2023time, zhou2024one, gruver2024large, zhang2023instruct, li2024frozen, jiang2023health, xue2024prompt, mizuno2022generation, zhao2023survey, wang2023would}. In addition, it has been shown that LLMs embed spatial and geographical knowledge and that this knowledge can be effectively extracted \citep{manvi2024geollm, tan2023promises}, making it plausible to use LLMs for geographic- and mobility-related tasks. Concerning the field of human mobility, LLMs have been successfully used to generate individual-level trajectories \citep{mizuno2022generation} and predict mobility demand at a Point of Interest (POI) level \citep{xue2022leveraging,xue2024prompt}. 



In our work, we have performed an extensive evaluation of 15 different LLMs to understand if these models can act as next-location predictors. In particular, we tested 
Llama 2 7B, 13B and 70B \citep{touvron2023llama}, Llama 2 Chat 7B, 13B and 70B \citep{touvron2023llama}, Llama 3 8B and 70B \citep{dubey2024llama}, Llama 3 8B Instruct and 70B Instruct \cite{dubey2024llama}, Llama 3.1 8B, Mistral 7B, GPT-3.5 \citep{OpenAI_2022}, GPT-4 \citep{OpenAI_2022}, and GPT-4o \citep{OpenAI_2022}. We also tested eight additional LLMs that led to negative results (e.g., do not understand the task, do not provide answers). These models are Phi-1.5 \citep{textbooks2}, Phi-2 \citep{phi2}, Phi-3 \citep{phi3}, Gemma 2B \citep{gemmateam2024gemma}, GPT-J \citep{gpt-j}, Dolly 3B, Dolly 7B, Dolly 12B\citep{dolly}. Examples of the obtained negative results and outputs are presented in Supplementary Information 3. All the models are evaluated on three real-world mobility datasets. Two of them are open datasets collected in the cities of New York and Tokyo. We decided to add a private dataset collected in Ferrara, Italy, that we mainly used to ensure that results are not positively biased and there are no data contamination issues.

\begin{figure}[H]
    \centering
    \includegraphics[width=1\linewidth]{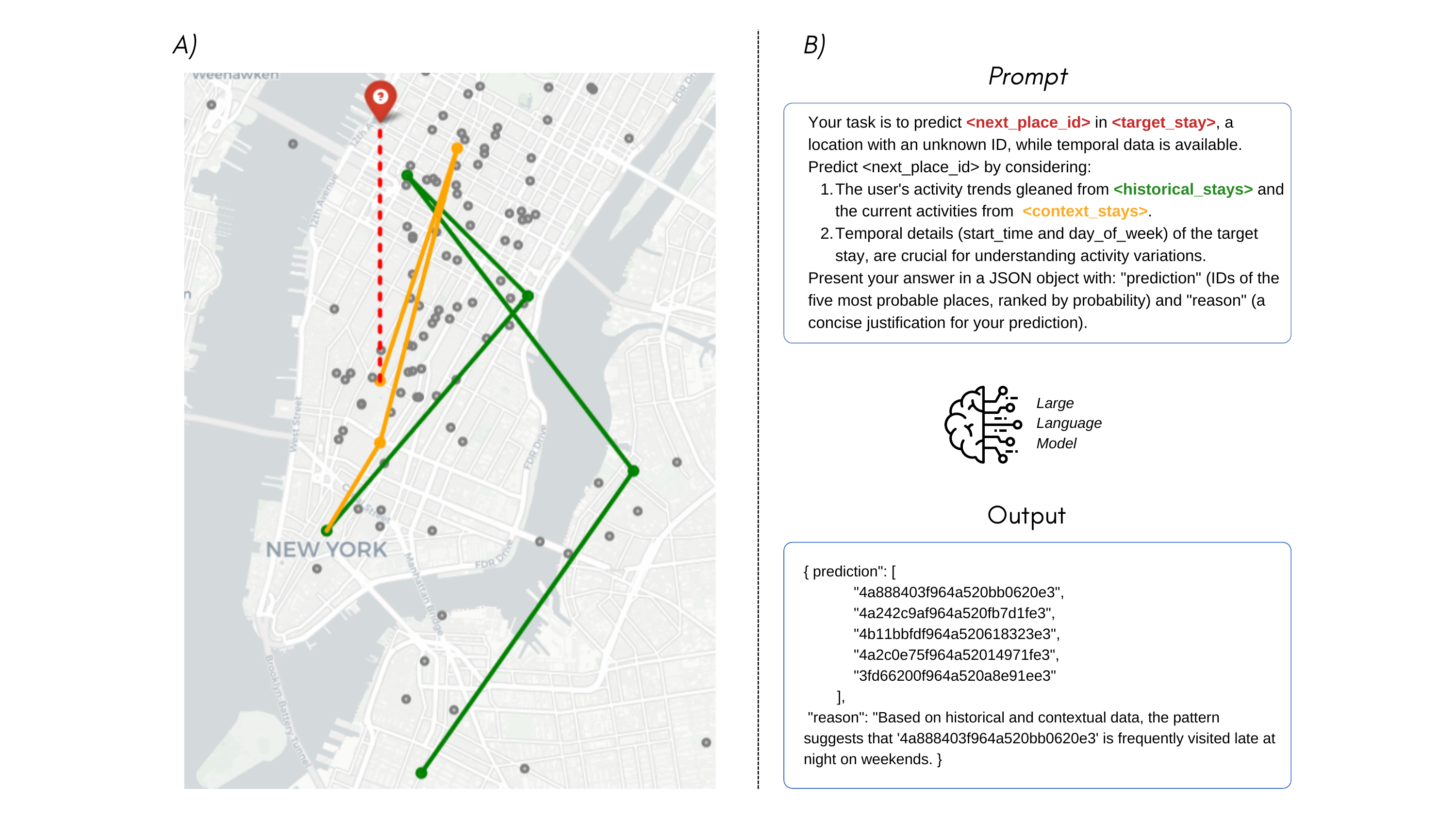}
    \caption{\textbf{Framework for predicting the next location using LLMs.} In panel \textit{A)}, we graphically describe the problem of next-location prediction. Given a set of historical (green) and contextual (yellow) user visits, a next location predictor is a model that predicts the next place (red) they will visit. In panel \textit{B)}, an example of how we query an LLM and an example of response.
    }
    \label{fig:framework}
\end{figure}

After showing the geographic transferability issues of some DL models commonly used as baselines for NL, we tested the zero-shot prediction capabilities of these LLMs in the same context. Remarkably, our findings indicate that LLMs can effectively operate as zero-shot NLs, achieving accuracies up to 36.2\%. Our framework is graphically represented in Figure \ref{fig:framework} where in panel \textit{A)}, we represent the task and in panel \textit{B)}, we showcase how LLMs are prompted and we provide an example of response. 

We also explored the role of in-context learning (ICL) testing the LLMs with examples of input and expected output \citep{brown2020language, liu2023pre, wei2022chain}. In particular, we systematically analyzed the impact of zero-shot, one-shot and few-shot prompting on LLMs' performances, finding that one-shot and few-shot prompts do not imply better performances.
Further, we examined the influence of prompted historical information, such as the number of past location visits of a user and the context of the current trajectory. We observed that lowering the available information can have severe performance drops (up to a relative drop of -58.20\%). On the other hand, providing additional historical visits can improve performances up to 16.18\% in terms of relative improvement. 

Finally, we tested the models for data contamination \citep{sainz2023nlp}. As LLMs are trained on a vast amount of publicly available data, there is the risk of obtaining positively biased results because data were observed during training. To avoid issues related to data contamination, \emph{(i)} we tested the models for it, and \emph{(ii)} we ran all the experiments using two publicly available datasets \citep{yang2014modeling} and a private dataset \citep{bucchiarone2023play}. As a result, data contamination was not an issue, and the performances obtained using public datasets were comparable to the ones obtained using our private dataset. In addition to all the previous experiments, we also explored the possibility of using LLMs as text-based explainers for NL, showing that LLMs can effectively provide a text-based explanation of why a specific location was shortlisted as a potential next location.  

\section*{Results}
\subsection{Next-Location Predictors are not Geographically Transferable}
\label{sec:traditional_zero}
Next-location prediction is the problem of predicting the next location an individual will visit, and a formal problem definition is available in Methods. As a first step, we show that the geographic transferability of next location predictors is actually a problem. Although a definitive benchmark for identifying state-of-the-art models in NL  is lacking \citep{yabe2024enhancing}, we selected six widely adopted DL baselines for our analysis, as detailed in the Methods section. We show that once models are trained in a city, the deployment in different areas (e.g., another city) leads to severe performance drops due to known non-transferability and limited generalization capabilities issues \citep{luca2023trajectory, bontorin2024mixing}. In particular, we train and test the models on two publicly available datasets collected in New York and Tokyo \citep{yang2014modeling}. The two datasets are detailed in Methods. Here, we are using only the datasets of Tokyo and New York to avoid having different data sources, which may lead to different self-selection biases and spatial granularity. Thus, results may be less interpretable \citep{luca2020survey, barbosa2018human}.
To carry out our analysis, we first split each dataset into a training set (70\%), a validation set (10\%), and a test set (20\%) following the commonly adopted user-based approach proposed in \citep{feng2018deepmove}. This implies that after segmenting the raw dataset into trajectories, for each user, we select their first 70\% of trajectories as training, the following 10\% as validation and the last 20\% as the test ones.
We trained the models using the training set partitions from both New York and Tokyo using the hyperparameters listed in the Supplementary Information 2 file. Subsequently, we evaluated each trained model's performance on the alternate city's test set. As an evaluation metric, we use Accuracy@$k$ (ACC@k), focusing in particular on Accuracy@5 (ACC@5). The formal definition of the evaluation metrics is reported in the Methods section. In general, Accuracy@$k$ can be seen as a score between 0 and 1, representing how many times the actual next location was among the likely $k$ locations predicted by a model. ACC@5 scores for the selected DL baselines are reported in Table \ref{tab:traditional_dl_transfer}.

\begin{table}[H]
\centering
\resizebox{0.4\textwidth}{!}{
\begin{tabular}{lcc}
\toprule
         & \begin{tabular}[c]{@{}c@{}}\textbf{Train:} {New York}\\ \textbf{Test:} {Tokyo}\end{tabular} & \begin{tabular}[c]{@{}c@{}}\textbf{Train:} {Tokyo}\\ \textbf{Test:} {New York}\end{tabular} \\ \midrule
RNN      & 0.028                                 & 0.029                                 \\ 
ST-RNN   & 0.031                                 & 0.038                                 \\ 
DeepMove & 0.031                                 & 0.046                                 \\ 
LSTPM    & 0.030                                 & 0.046                                 \\ 
STAN     & 0.038                                 & 0.048                                 \\ 
MobTCast & \textbf{0.042}                        & \textbf{0.049}                        \\ \bottomrule
\end{tabular}
}
\caption{\textbf{ACC@5 values of the traditional DL models when trained using data for New York/Tokyo and tested on Tokyo/New York}. MobTCast is the best-performing model in both cases, with an ACC@5 of 0.042 when trained in New York and tested in Tokyo and 0.049 when trained in Tokyo and tested in New York.}
\label{tab:traditional_dl_transfer}
\end{table}

As we can see, MobTCast obtained the best results with a 0.042 ACC@5 when trained in New York and tested in Tokyo. When trained in Tokyo and tested in New York, MobTCast reached an ACC@5 of 0.049. In this scenario, LSTPM, DeepMove and STAN reached comparable ACC@5 values of 0.046, 0.046 and 0.048, respectively. These results highlight severe limitations regarding generalization power and geographic transferability of widely adopted DL models \citep{luca2023trajectory, luca2020survey}. Such limitations also prevent us from using these models as zero-shot predictors effectively. Designing a zero-shot NL remains an open challenge \citep{luca2020survey}, and in this paper, we test whether LLMs can represent a solution for language-based zero-shot next-location predictions. 

\subsection{Language-Based Zero-Shot Next Location Predictors}
\label{sec:llm_zero}
Before showcasing the results obtained by the LLMs and describing the experiments, it is important to mention that the results we will discuss in the main text concern the city of New York. We selected New York as an example as we obtained the same trends and conclusions for Tokyo and Ferrara. The results for Tokyo and Ferrara are presented in Supplementary Information 6-10. To ensure the robustness of the findings, the accuracies, along with their standard deviations, were obtained from three independent runs of each model on each dataset maintaining the same test set.
To test LLMs' capabilities to act as zero-shot NLs, we designed a structured prompt tailored to provide LLMs with the same information commonly provided to NLs \citep{luca2020survey}. Specifically, we prompted the LLMs with data on individuals' historical and contextual visits, extracted according to the formal problem definition described in the Methods section, and we asked the LLMs to predict the top five possible next locations an individual will visit. Moreover, to enhance the comparability among the LLMs, we choose to use the same prompt for all the selected LLMs. The list of LLMs and their description are in the Methods section. Examples of the adopted prompts can be observed in Figure \ref{fig:framework} and more extensively in Supplementary Information 3. We started by specifying the task, data and desired output, and then we provided a list of spatio-temporal historical points and contextual visits to ask the model for a prediction eventually. 

\begin{figure}[H]
    \centering
    \includegraphics[width=1\linewidth]{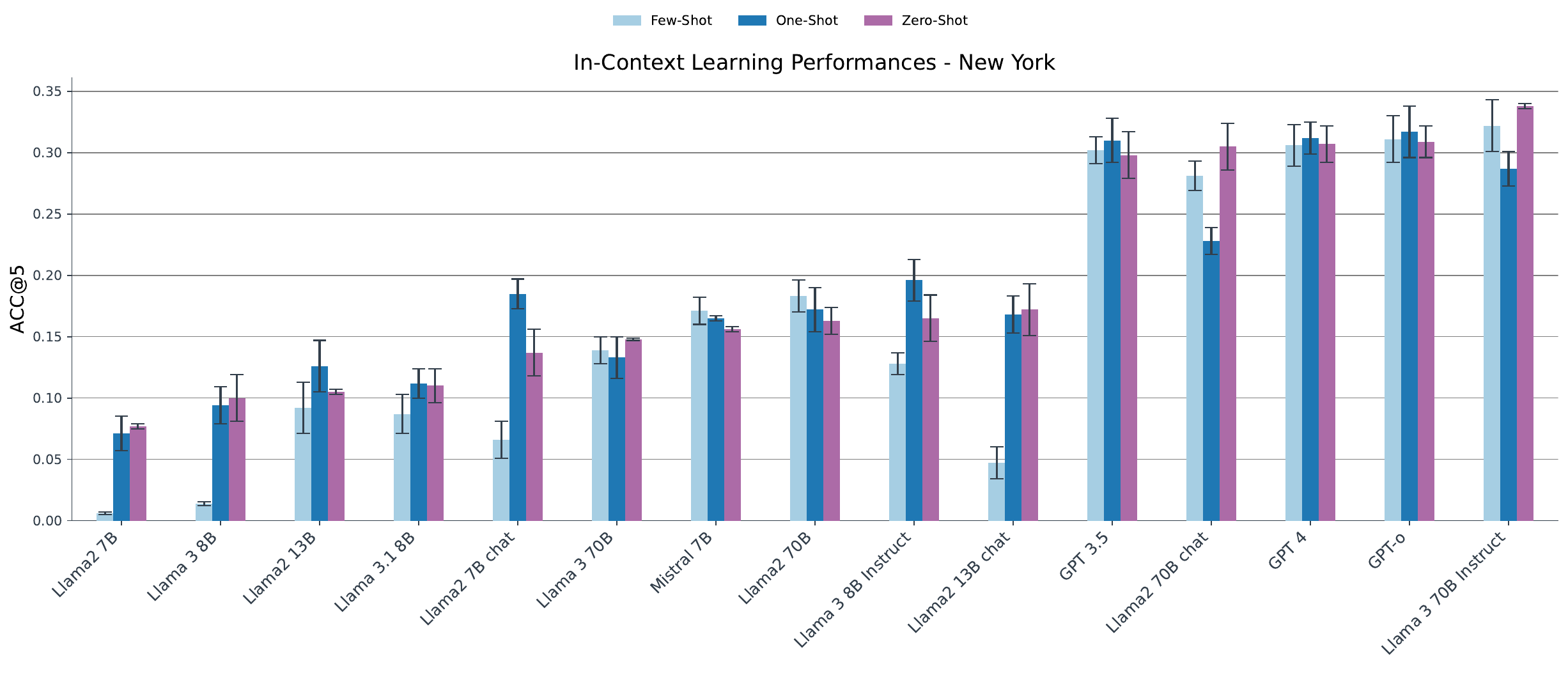}
    \caption{\textbf{ACC@5 values of LLMs} with zero-shot (purple), one-shot (dark blue) and few-shot (light blue) prompts in New York. The other cities have similar results and are reported in Supplementary Information 7 and 9 files. Overall, the best performances are reached by larger models regardless of the prompt adopted. In particular, Llama 3 70B Instruct is the best-performing model, with GPT-4o, GPT-4, Llama 2 Chat 70B  and GPT-3.5 reaching similar performances. Smaller models, especially non-instruct or non-chat ones, have lower performances. Performances may vary significantly depending on the prompt adopted.
    }
    \label{fig:acc_zero_shot}
\end{figure}


In Supplementary Information 7 and 9, we present the ACC@5 results of the LLMs across three datasets. The results for New York, illustrated in Figure \ref{fig:acc_zero_shot}, show that Llama 3 70B Instruct achieved the highest performance in zero-shot, reaching an ACC@5 of 0.338. Notably, also GPT-4 and GPT-4o performed well, with ACC@5 scores of 0.307 and 0.309, respectively.
The figure provides a clear comparison of all the models' performances across different prompting strategies, with the zero-shot performance highlighted by the purple right-most bar for each model Turning our attention to Tokyo, Llama 3 70B Instruct maintains its superior performance with an ACC@5 of 0.341, recording the best performance across all the datasets.
In Ferrara, the performance shifts slightly, with  GPT-4o and Llama 3 70B Instruct reaching very close performances of 0.288 and 0.287, respectively.

As we may expect, we find that bigger models perform better than the smaller ones in zero-shot scenarios \citep{kaplan2020scaling}. 
Interestingly, we also find some exceptions: for example, Mistral 7B consistently outperforms Llama 2 13B models despite having fewer parameters, and Llama 3 8B Instruct achieves performances that are comparable to Llama 2 13B in its Chat version.
In Supplementary Information 5, 6 and 7, we also report the results for ACC@1, ACC@3 and ACC@5, showing that the insights obtained are consistent across all the evaluation metrics. 
Remarkably, we also observed that LLMs performed significantly better than DL state-of-the-art next-location predictors when tested in a setting that was as close as possible to zero-shot NL. In particular, using GPT-4o with additional historical information led to a relative improvement of 638.77\% in New York and 759.52\% in Tokyo.
Detailed results for all the models and evaluation metrics are reported in Supplementary Information 5, 6 and 7.

In addition to zero-shot prompting, we also tested one- and few-shot prompting, as LLMs may benefit from so-called In-Context Learning.

\subsection{In-Context Learning with One-Shot and Few-Shot Performance Evaluation}
\label{sec:llm_few}
 In-Context Learning (ICL) involves providing LLMs with examples of input and expected outputs so that models may be facilitated in understanding the task. In this sense, one-shot and few-shot prompting involve providing the LLMs with a prompt that includes one or a few examples of pairs of input and expected output instead of directly asking the models to solve a task (zero-shot prompting). 
 
Here, we explored how different predictive prompting strategies affect the accuracy of the selected models. Results are reported in 
Supplementary Information 9 and 10 and can be visually observed for the city of New York in Figure \ref{fig:acc_zero_shot} where the left-most light blue bars show results for few shot prompts and the middle dark blue bars highlight accuracies for one shot prompts. As trends and conclusions are the same for Tokyo and Ferrara, the relative plots are available in the Supplementary Information 9 and 10 files.
GPT-4o is the model that achieved the best performance regardless of the dataset and prompting technique. Also, it reached higher accuracies using one-shot prompts. Likewise, Llama 2 13B and Llama 2 Chat 7B obtained the best performances with a one-shot prompting strategy with performances comparable to zero-shot ones. In general, Llama models tend to achieve their best performances with zero-shot or one-shot prompts. Finally, Mistral 7B gives its best with few-shot prompts. The role of prompting techniques is fundamental and there are instances (e.g., Llama 2 13B Chat) in which the performance relative drop between the different prompts can be up to 77.2\%. In particular, our analysis indicates that smaller models, especially within the Llama family, often struggle in few-shot scenarios. These models, despite showing adaptability in zero-shot and one-shot scenarios, tend to underperform when prompted with multiple examples. In contrast, this trend is less pronounced among larger models,  which regardless of the family models keep the performances comparable across the prompts.

\subsection{The Role of Contextual and Historical Visits}
\label{sec:hist_context}
In our prompts, we provided LLMs with 15 historical visits and 6 contextual visits. We modified these numbers of historical visits ($H$) and contextual visits ($C$) to carry out an ablation study. We analyzed how accuracies change by halving and doubling the number of $C$ and $H$. Also, we explored how performances are impacted when no $C$ or $H$ are provided.

\begin{figure}[H]
    \centering
    \includegraphics[width=1\linewidth]{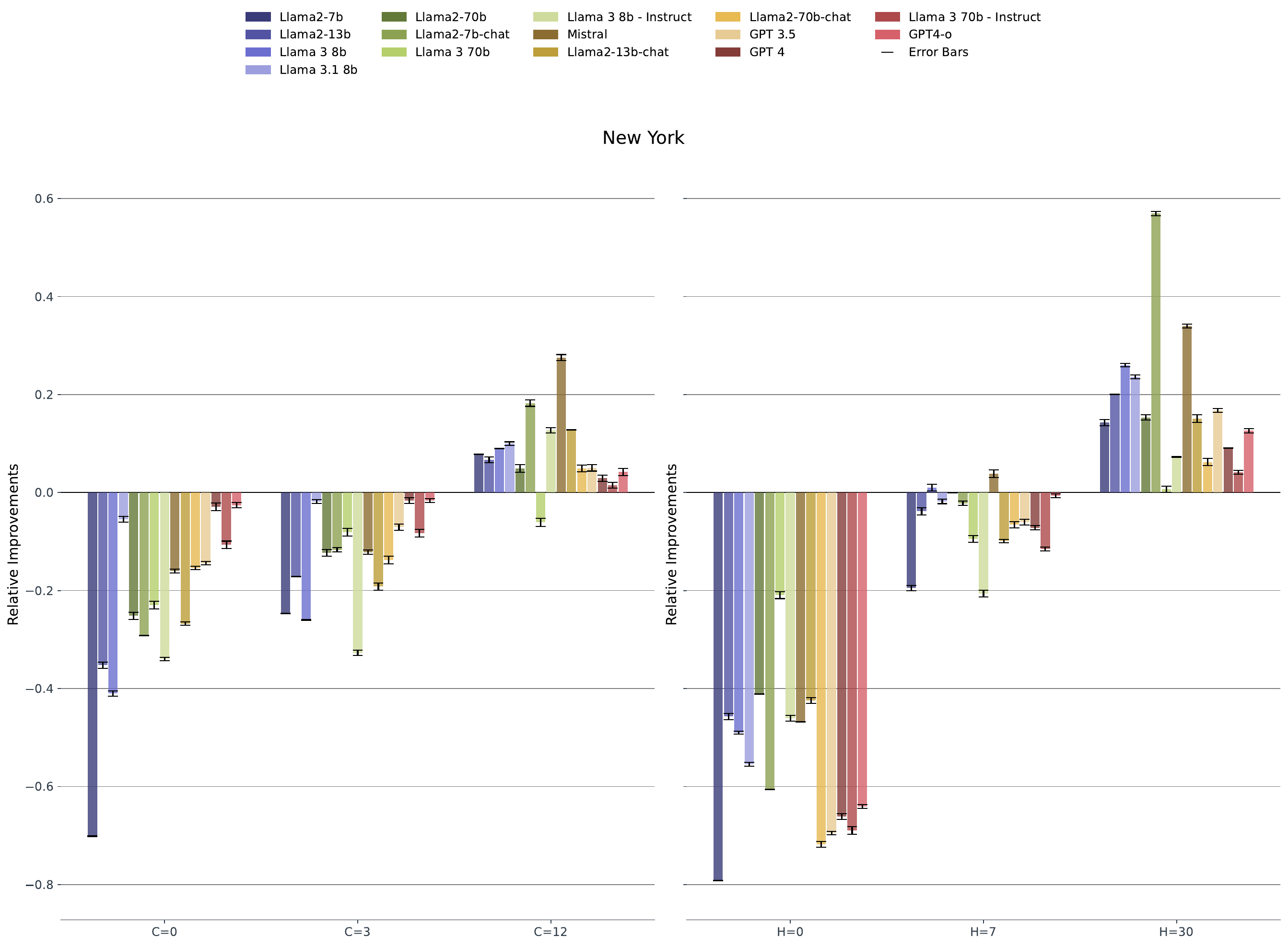}
    \caption{ \textbf{Results in terms of relative improvements and drops} with respect to the scenario with $C=6, H=15$ for the city of New York when we modify the availability of contextual $C$ (left) and historical $H$ (right) information. Adding locations to $H$ or $C$ ($C=12$ or $H=30$) led to an improvement of ACC@5 while limiting such information led to a drop of performances that can be more ($C=0$ or $H=0$) or less ($C=3$ or $H=7$) severe.}
    \label{fig:conf_c_h}
\end{figure}

In Figure \ref{fig:conf_c_h}, we showcase results in terms of relative improvements and drops with respect to the scenario with $C=6, H=15$ for the city of New York. We can observe similar trends for Tokyo and Ferrara, which are shown in Supplementary Information 8 and 10 files. 
As we can also see in Figure \ref{fig:conf_c_h}, increasing the number of visits in $C$ or $H$ leads to an improvement in terms of ACC@5 regardless of the model while removing or limiting such information led to a drop of performances. With the default prompt ($C=6; H=12$), ACC@5 varies between 0.077 
obtained by Llama2 7B and 0.338 
obtained by Llama 3 70B Instruct 
. The average ACC@5 in New York is 0.193
When we double contextual information ($C=12$), the new ACC@5 varies between 0.083-0.343 
with an average performance of 0.206 
corresponding to an average relative improvement of 6.92\% 
When we double the number of historical visits ($H=30$), ACC@5 ranges between 0.088-0.352 (New York)
with an average performance of 0.221 
in terms of ACC@5 and an average relative improvement of 14.84\% 
. We also tested prompts with a reduced number of visits (i.e., $C=3$ or $H=7$) or with no contextual ($C=0$) or historical ($H=0$) information at all. When we use half of the contextual information ($C=3$), we experienced a reduction of the ACC@5 ranges between 0.058-0.310 
with an average ACC@5 of 0.171 
and an average relative drop of -11.07\% 
. Concerning the reduction of historical information ($H=7$), the average performance is 0.180 
with accuracies varying between 0.062-0.307 
, corresponding to an average relative drop of -6.33\%.
The extreme case in which we completely remove contextual or historical information leads to severe drops in performances. With $C=0$, relative drops are of -18.09\% and ACC@5 of 0.053-0.158  
Also, by setting $H=0$, we obtained average relative drops of -58.20\% with a corresponding ACC@5 of 0.016-0.081.


We also tried to quantify the importance of $C$ and $H$ in the standard prompt (i.e., $C=6; H=15$). Given all the correctly predicted next locations, we look at the number of instances in which the next location was part of $H$ and the percentage of times in which the prediction was based on $C$. We found that, for the dataset of  New York, in the 55.88\% of the cases it was possible to find the next location mentioned both in $C$ and $H$. 
Similarly, in 31.01\% of the cases, 
the next location was only available in historical visits $H$. Eventually, in 13.11\% of the trajectories,
the actual next location was only present in $C$. This analysis suggests why changing the number of historical visits $H$ is more impactful than changing the number of contextual visits $C$.

\subsection{Searching for Data Contamination}
\label{sec:contamination}
An important analysis when it comes to LLMs is to understand how much the results and accuracies obtained are biased as open datasets may have already been observed during training.  New York and  Tokyo datasets are publicly available with many instances on websites and repositories (e.g., GitHub codes). Thus, we must be sure that the performances do not depend on potential data contamination issues. To mitigate this problem, we have done the following three actions.

\noindent \textbf{Test on a private dataset.} As a first counteraction, we analyzed the performances also on a private dataset (Ferrara dataset) which we are sure was not used to train any LLM. 
Results are presented in Supplementary Information 5, 6 and 7 files.
The performances we obtained are similar to the ones on other publicly available datasets. Thus, LLMs can be effectively used as zero-shot NLs even on datasets that we can guarantee have not been observed in any way during the training phase.

\textbf{Carry out a data contamination analysis \citep{sainz2023nlp}.} We designed a quiz (following \citep{golchin2023data}) where we provided the LLMs with precise questions about the content (e.g., rows) of the datasets and four possible answers. Only one of the answers consists of entries coming from the same dataset, while the other options were designed following the suggestions in \citep{golchin2023data}. These suggestions, involve designing plausible but incorrect choices that closely mimic the truth in style and substance without actually being correct. This method helps in distinguishing whether LLMs are merely outputting learned information or genuinely understanding and processing the input to generate informed responses. All the models could not select the correct option, highlighting that the results obtained are not biased by data contamination. Examples of the quiz questions are provided in the Supplementary Information 11 file.

\textbf{Analyze unusual outputs.} As a last analysis, we also carefully examined specific outputs of the models. What we found is that the models provided three types of outputs: \emph{i)} empty outputs and no next-location; \emph{ii)} correct predictions; \emph{iii)} hallucinated location identifiers. Concerning the latter, we selected all the answers with a location identifier that was not included in the historical or contextual visits and we checked the outputs. We found that the structure of the identifiers was similar to real Foursquare location identifiers  (e.g., comparable numbers of characters) but none of them correspond to existing locations. The Supplementary Information 3 file gives an example of this hallucinated output.

\subsection{Large Language Models Can Explain their Predictions}

We also designed our prompts to obtain an explanation about what drove the LLMs to put a specific next location. In Table \ref{tab:expl_example}, we report an example of an explanation provided by GPT-4o. We noticed that the level of detail of the explanation mainly depends on the number of parameters of the models, with larger models being able to refer to specific location identifiers and smaller models that are capable of providing reasonable but more general explanations. An example of an explanation for each model is provided in the Supplementary Information 4 file.
Remarkably, regardless of models' sizes, the explanations focus on locations visited in similar circumstances (e.g., weekends versus weekdays, similar hours). In accordance with multiple studies about NL \citep{luca2020survey,schneider2013unravelling,song2010limits}, focusing on previously visited locations may represent an effective way to perform zero-shot next location prediction and it may justify the good performances obtained by LLMs when used as zero-shot NLs.

\begin{table}[h]
\resizebox{0.8\textwidth}{!}{%
\begin{tabular}{@{}ll@{}}
\toprule
\multicolumn{2}{l}{\begin{tabular}[c]{@{}l@{}}\small \textbf{Explanation GPT-4o:}\\ \\ The user has frequently visited the places '4d48864b551db1f722e9a330', \\ '4a5dd98ef964a520c2bd1fe3', '3fd66200f964a520def11ee3', '42ec1480f964a5209e261fe3', and \\ '4a3d735cf964a5204ba21fe3'. In addition, context locations provide recent activities at places \\ '4a5dd98ef964a520c2bd1fe3', '3fd66200f964a520def11ee3', and '4a3d735cf964a5204ba21fe3'. \\ It is likely that the user will visit one of these places again.\end{tabular}}          \\ \bottomrule

\end{tabular}
}
\caption{An example of an explanation provided by GPT-4o. The model identifies frequently visited locations by the user and leverages recent activity data to predict NL. }
\label{tab:expl_example}
\end{table}

\section*{Discussion}
In this paper, we tested the capabilities of GPT-4o, GPT-4, GPT-3.5, Llama 2, Llama 3, Llama 3.1 and Mistral to act as zero-shot next-location predictors. The task consists of predicting the next location an individual will visit, given a set of historical and contextual visits. 
While there are many sophisticated DL models specifically designed to solve this task, having models that are geographically transferable and that can effectively predict future whereabouts in data-scarce scenarios remains an open challenge. In our work, we show that, indeed, LLMs can act as zero-shot location predictors obtaining remarkable results (i.e., up to 0.362 - ACC@5) corresponding to relative improvements up to 638\% with respect to DL-based next-location predictors when tested in a scenario as close as possible to the zero-shot. We also tested Gemma, Phi-1.5, Phi-2, Phi-3, GPT-J and Dolly (3B,7B,12B) but we found that, differently from other LLMs, such models tend to misunderstand the task providing empty or unusable answers. We also explore the role of in-context learning and how it affects performances, showing that the number of examples of input/output pairs provided to the models has mixed effects on performances, with some models obtaining their best performances with few-shot prompts and others with one- or zero-shot prompts. Moreover, we explored the role of the amount of available historical and contextual information. We found that performances may vary between -26.38\% and 56.92\% (relative improvements), with data-richer prompts leading to better performances, with the historical visits more impactful in the performance. 
Finally, we investigated the possibility of using LLMs as text-based explainers for next-location prediction. We showed that LLMs can explain why the model selected that particular location. 
In our study, we tested a significant number of LLMs using a single prompt to ensure comparable results across the different models. However, each LLM comes with model-specific knowledge and prompt formulations. To maximize an LLM's performance, prompts should be tailored to each model.
However, beyond ensuring comparability, using a single prompt helped us manage costs. Indeed, testing LLMs incurred nearly 1,000 USD in API call expenses and running the experiments on local machines may lead to similar or higher management costs. 
We also found that not all the tested LLMs could understand the task and provide meaningful answers. Also, the working ones could hallucinate outputs (i.e., generate plausible but non-existent location identifiers). It would be interesting to investigate if allowing LLMs to access external resources (e.g., list of location identifiers, business type) can improve performances and reduce hallucinations. Also, improving performances may allow policymakers to deploy LLMs in real-world use cases. While an improvement of more than 600\% in terms of ACC@5 is remarkable, an ACC@5 of 34\% may not be sufficiently reliable for critical use cases (e.g., natural disasters).  
From a societal perspective, the use of LLMs in predicting human mobility can significantly impact urban planning and public policy. Accurate predictions can help to improve public services like transportation and infrastructure development. They can also enhance responses to emergencies and disasters by optimizing evacuation routes and resource allocation. In this sense, LLMs' capabilities to effectively perform next-location prediction in zero-shot and with a small amount of information may be a critical factor. %
On the other hand, the use of LLMs also raises important ethical and societal concerns. A critical ethical issue is the potential for bias in the models \citep{manvi2024geollm}. LLMs trained on large datasets may inherit biases present in their training data, which could generate predictions based on demographic or geographical factors, leading to discriminatory and unfair outcomes. Different demographic groups may have varying mobility patterns that are not equally represented in the training datasets, leading to less accurate predictions for these groups and potentially exacerbating existing inequalities.


\section*{Methods}

\subsection{Task Definition}
\label{sec:task_definition}
Next-location prediction is commonly defined as the problem of predicting the next location an individual will visit given their historical movements, typically represented as spatio-temporal trajectories \citep{luca2020survey}.

\begin{definition}[Trajectory]
A spatio-temporal point $p=(t, l)$ is a tuple where $t$ indicates a timestamp and $l$ is a geographic location. 
A trajectory $P = p_1,p_2,\dots,p_n$ is a time-ordered sequence of $n$ spatio-temporal points visited by an individual, who may have several trajectories, $P_{1}, \dots, P_{k}$, where all the {\color{black} locations in $P_i$ are }visited before locations in $P_{i+1}$.
\end{definition}

We further filtered trajectories following the methodology proposed in DeepMove \citep{feng2018deepmove}. This involves filtering out users with fewer than 10 records. Also, we selected a 72-hour interval as the threshold for distinguishing between separate trajectories. Subsequently, any user with fewer than five trajectories is excluded from the analysis.

{Each trajectory $P$ of a user is composed of Historical and Contextual visits. We identified Historical visits $\mathcal{H}$ and Contextual visits $\mathcal{C}$ as follows.}

\begin{definition}[Historical visits] 
Historical visits $\mathcal{H} = \{h_1, h_2, \ldots, h_n\} $ are the $n$ spatio-temporal points that represent the user's long-term mobility patterns and are visited immediately before the Contextual visits $ \mathcal{C} $. Formally, 
    $$ \mathcal{H} = \{h_i \mid h_i \in P_k \text{ and } h_i \prec c_1 \} $$
where $ c_1 = \min(\mathcal{C}) $ represents the earliest point in the Contextual visits. 
\end{definition}

\begin{definition}[Contextual visits] 
Contextual visits $ \mathcal{C} = \{c_1, c_2, \ldots, c_m\} $ are the $m$ spatio-temporal points that capture the user's short-term mobility patterns. The last point of the Contextual visits, $c_m$, is the same as the last point of the current trajectory, $c_m = p_n$. Thus, these set of points are visited immediately before the target point $ p_{n+1} $. Formally, 
    $$ \mathcal{C} = \{c_j \mid c_j \in P_k \text{ and } c_j \prec p_{n+1} \} $$
\end{definition}


\noindent We formalize the problem of next-location prediction as follows.


Given:
\begin{itemize}
    \item The current trajectory of an individual $ P_k = \{p_1, p_2, \dots, p_n\} $, with at least two points. 
    \item The historical trajectories of the individual $ \mathcal{L} = \{P_1, P_2, \dots, P_{k-1}\} $.
\end{itemize}
Next-location prediction is the problem of forecasting the next location point $p_{n+1} \in P_{k}$. 

In other words, a next-location predictor (NL) is a function $$ \mathcal{M}: (\mathcal{H}, \mathcal{C}) \rightarrow p_{n+1} $$
where $ \mathcal{H} $ are the Historical visits formed by filtering the earlier parts of $ P_k $ and possibly earlier trajectories $ P_1, \dots, P_{k-1} $ and $ \mathcal{C} $ are the Contextual visits defined as the last few points of $ P_k $ leading up to $ p_n $.

\subsection{Datasets}
We leveraged two datasets collected on Foursquare, a location-based social network where users can check-in at POIs. Data were collected in New York and Tokyo \citep{yang2014modeling}. Each entry consists of a user identifier, a location identifier, geographical coordinates, a timestamp, and a venue's category. In New York, we have 4,390 users and 13,960 unique POIs, corresponding to 12,519 distinct trajectories. In Tokyo, we have 935 users, 21,394 unique POIs, and 34,662 trajectories.

We also employed a private dataset to mitigate potential data contamination issues. It consists of GPS trajectories produced by cyclists in Ferrara, a Northern Italian city of 132,000 inhabitants \citep{bucchiarone2023play}.
We divided the city into 200 x 200 square meters and ended up with 2,488 unique locations. We then computed the stop locations \citep{aslak2020infostop} (see Supplementary Information 1 for a more detailed explanation). We selected 3,284 trajectories belonging to 127 users. Table \ref{tab:data_summary} provides a concise summary of key statistics for all three datasets.

\begin{table}[H]
\centering
\small
\resizebox{0.5\textwidth}{!}{%
\begin{tabular}{llcccc}
\hline
&                              & \textbf{Users} & \textbf{Locations} & \textbf{Trajectories} & \textbf{Time Span} \\ \hline
FSQ New York     & \citep{yang2014modeling}       & 4,390  & 13,960    & 12,519 & 8 Months                \\
FSQ Tokyo   & \citep{yang2014modeling}       & 935   & 21,394    & 34,662 & 8 Months                \\
Ferrara &   \citep{bucchiarone2023play}   & 127    & 2,488      & 3284    & 3 Months                \\ \hline
\end{tabular}%
}
\caption{Number of users, unique locations, and trajectories for each dataset.}
\label{tab:data_summary}
\end{table}

\subsection{Evaluation Metrics}
We employed the $k$-accuracy (ACC@$k$) metric, a standard in the field of NL \citep{luca2020survey}. 
This metric quantitatively measures the frequency with which the actual next location a person visits appears within the top $k$ predictions provided by a model. 
In our analyses, we primarily focused on ACC@$1$, ACC@$3$ and ACC@$5$. Results concerning ACC@$5$ are presented in the main paper while ACC@$1$ and ACC@$3$ are presented in Supplementary Information.

$$\text{ACC@}k = \frac{1}{N} \sum_{i=1}^N \mathbf{1}(\text{rank} (p_{n+1}) \leq k) $$

\subsection{Baselines}
Our first goal is to show that current state-of-the-art models are not zero-shot predictors and are not geographically transferable. To do so, we selected the following widely adopted models as baselines:
\textbf{RNNs} \citep{hopfield1982neural} are widely used as building blocks of NLs to capture spatial and temporal patterns in the trajectories; 
\textbf{ST-RNNs} \citep{liu2016predicting} add to standard RNNs time-specific and space-specific transition matrices; 
\textbf{DeepMove}  \citep{feng2018deepmove} leverages attention mechanisms to capture spatio-temporal patterns in individual-level historical trajectories;
\textbf{LSTPM} \citep{sun2020go} combines long- and short-term sequential models, where long-term patterns are modelled using a non-local network \citep{wang2018non}, while short-term preferences are captured using a geographic-augmented version of the concept of dilated RNNs \citep{chang2017dilated};
\textbf{STAN} \citep{luo2021stan} captures spatio-temporal data by employing a multi-modal embedding to depict trajectories and a spatio-temporal attention mechanism to discern patterns within the data; and finally
\textbf{MobTCast} \citep{xue2021mobtcast} incorporates a transformer encoder-based structure to predict the next POI, taking into account temporal, semantic, social, and geographical contexts. 
All the baselines, with the exception of MobTCast, are available in LibCity \citep{libcity}; while MobTCast implementation can be found in \texttt{https://github.com/xuehaouwa/POI-TForecast}
\subsection{Large Language Models}
Concerning LLMs, we extensively evaluated the following models. First, we examined the \textbf{Llama series}, including \textbf{Llama 3.1} (8B) and \textbf{Llama 3} (7B, 70B), alongside their respective Instruct versions \citep{dubey2024llama}. Additionally, we evaluated the \textbf{Llama 2} models (7B, 13B, and 70B), as well as their Chat versions \citep{touvron2023llama}. We utilized Replicate APIs\footnote{https://replicate.com/} to test all Llama series, enabling a comprehensive analysis of their capabilities.
In parallel, we assessed models from the \textbf{GPT series} developed by OpenAI. This included \textbf{GPT-4}, \textbf{GPT-4o}, and \textbf{GPT-3.5} \citep{OpenAI_2022}, which were evaluated by querying the models directly through the OpenAI API\footnote{https://platform.openai.com/docs/api-reference/chat}. 
In addition, we also evaluated \textbf{Mistral 7B}  \citep{jiang2023mistral}, a 7 billion parameter model able to perform better than bigger models (e.g., Llama 2 13B) on many benchmarks. We also tested other models like Phi-1.5, Phi-2 \citep{phi2}, Phi-3 \citep{phi3}, Gemma 2B \citep{gemmateam2024gemma}, GPT-J \citep{gpt-j}, Dolly (3B, 7B, 12B) \citep{dolly}. The outputs from these models, however, were empty or contained text that did not represent locations, location identifiers or reasoning about possible next locations. Examples of prompts with relative outputs are provided in Supplementary Information 3.

\subsection*{Reporting summary}
Further information on research design is available in the Nature
Portfolio Reporting Summary linked to this article.

\begin{acknowledgements}
The work of B.L. and M.L. has been supported by the PNRR ICSC National Research Centre for High Performance Computing, Big Data and Quantum Computing (CN00000013), under the NRRP MUR program funded by the NextGenerationEU. B.L. also acknowledges the support of the PNRR project FAIR - Future AI Research (PE00000013), under the NRRP MUR program funded by the NextGenerationEU. B.L and M.L. have been supported by the European Union’s Horizon Europe research and innovation program under grant agreement No. 101120237 (ELIAS).
\end{acknowledgements}

\subsection*{Contributions}
C.B. processed the data. C.B. and M.L. performed the experiments. C.B. and M.L. designed the study. C.B., M.L. and B.L. contributed to interpreting the results and writing the paper.

\subsection*{Competing Financial Interests}
The authors declare no competing financial interests

\subsection*{Data Availability}
Foursquare data for New York City and Tokyo are available at \href{https://sites.google.com/site/yangdingqi/home/foursquare-dataset}{https://sites.google.com/site/yangdingqi/home/foursquare-dataset} as indicated in \citep{yang2014modeling}. Ferrara's dataset is a private dataset we are not authorized to release. Contact the authors for additional details.

\subsection*{Code Availability}
The code is available on GitHub at \href{https://github.com/ssai-trento/LLM-zero-shot-NL}{https://github.com/ssai-trento/LLM-zero-shot-NL}

\bibliography{biblio}

\renewcommand{\appendixname}{}

\appendix
\renewcommand{\thesection}{Supplementary Information \arabic{section}} 
\renewcommand{\thefigure}{SI~\arabic{figure}}
\setcounter{figure}{0}
\renewcommand{\thetable}{SI~\arabic{table}}
\setcounter{table}{0}

\section{Stop Location Computation}
\label{app:stoploc}
Given a dense trajectory, to compute the stop locations we identified each temporal sequence of GPS coordinates within a 65-meter radius, where a user stayed for a minimum of 5 minutes \citep{hariharan2004project}. Subsequently, we applied the DBSCAN algorithm \citep{ester1996density} to identify dense clusters of points within a distance of $\epsilon = \Delta_s - 5$.
We define these dense clusters as stop locations.

\section{Models' Hyperparameters}
\label{app:hyper}

\begin{table}[h!]
\caption{Hyperparameters used for the baselines}
\centering
\resizebox{0.5\textwidth}{!}{%
\begin{tabular}{l|lll|}
         & Foursquare NYC                                                              & Foursquare Tokyo                                                            & Our                                                                        \\ \hline
RNN      & \begin{tabular}[c]{@{}l@{}}lr: 0.005\\ epochs: 200\\ batch: 32\end{tabular}  & \begin{tabular}[c]{@{}l@{}}lr: 0.005\\ epochs: 200\\ batch: 32\end{tabular}  & \begin{tabular}[c]{@{}l@{}}lr: 0.05\\ epochs: 75\\ batch: 16\end{tabular} \\ \hline
ST-RNN   & \begin{tabular}[c]{@{}l@{}}lr: 0.001\\ epochs: 200\\ batch: 32\end{tabular}  & \begin{tabular}[c]{@{}l@{}}lr: 0.001\\ epochs: 200\\ batch: 32\end{tabular}  & \begin{tabular}[c]{@{}l@{}}lr: 0.05\\ epochs: 75\\ batch: 16\end{tabular} \\ \hline
DeepMove & \begin{tabular}[c]{@{}l@{}}lr: 0.001\\ epochs: 200\\ batch: 32\end{tabular}  & \begin{tabular}[c]{@{}l@{}}lr: 0.001\\ epochs: 200\\ batch: 32\end{tabular}  & \begin{tabular}[c]{@{}l@{}}lr: 0.05\\ epochs: 75\\ batch: 16\end{tabular} \\ \hline
LSTPM    & \begin{tabular}[c]{@{}l@{}}lr: 0.0001\\ epochs: 200\\ batch: 32\end{tabular} & \begin{tabular}[c]{@{}l@{}}lr: 0.0001\\ epochs: 200\\ batch: 32\end{tabular} & \begin{tabular}[c]{@{}l@{}}lr: 0.05\\ epochs: 75\\ batch: 16\end{tabular} \\ \hline
STAN     & \begin{tabular}[c]{@{}l@{}}lr: 0.0001\\ epochs: 200\\ batch: 32\end{tabular} & \begin{tabular}[c]{@{}l@{}}lr: 0.0001\\ epochs: 200\\ batch: 32\end{tabular} & \begin{tabular}[c]{@{}l@{}}lr: 0.05\\ epochs: 75\\ batch: 16\end{tabular} \\ \hline
MobTCast     & \begin{tabular}[c]{@{}l@{}}lr: 0.0001\\ epochs: 220\\ batch: 32\end{tabular} & \begin{tabular}[c]{@{}l@{}}lr: 0.0001\\ epochs: 220\\ batch: 32\end{tabular} & \begin{tabular}[c]{@{}l@{}}lr: 0.005\\ epochs: 75\\ batch: 16\end{tabular} \\ \hline
\end{tabular}%
}
\label{tab:DL_hyperparameters}
\end{table}

\newpage
\begin{table}[h!]
\caption{Hyperparameters used for LLMs}
\centering
\resizebox{0.5\textwidth}{!}{%
\begin{tabular}{l|l}
\hline
\textbf{Model} & \textbf{Hyperparameters (All Datasets)} \\ \hline
\textbf{GPT-4o} & \begin{tabular}[c]{@{}l@{}}temperature: 0.01\\ max\_new\_tokens: 150\\ max\_tokens: 1000\end{tabular} \\ \hline
\textbf{GPT-4} & \begin{tabular}[c]{@{}l@{}}temperature: 0.01\\ max\_new\_tokens: 150\\ max\_tokens: 1000\end{tabular} \\ \hline
\textbf{GPT-3.5} & \begin{tabular}[c]{@{}l@{}}temperature: 0.01\\ max\_new\_tokens: 150\\ max\_tokens: 1000\end{tabular} \\ \hline
\textbf{Llama 3.1} & \begin{tabular}[c]{@{}l@{}}temperature: 0.01\\ max\_new\_tokens: 200\\ top\_p: 0.95\\ top\_k: 250\\ length\_penalty: 1\\ presence\_penalty: 0\end{tabular} \\ \hline
\textbf{Llama 3} & \begin{tabular}[c]{@{}l@{}} temperature: 0.01\\ max\_new\_tokens: 200\\ top\_p: 0.95\\ top\_k: 250\\ length\_penalty: 1\\ presence\_penalty: 0\end{tabular} \\ \hline
\textbf{Llama 3 Instruct} & \begin{tabular}[c]{@{}l@{}}temperature: 0.01\\ max\_new\_tokens: 200\end{tabular} \\ \hline
\textbf{Llama 2 Chat} & \begin{tabular}[c]{@{}l@{}}temperature: 0.01\\ max\_new\_tokens: 200\end{tabular} \\ \hline
\textbf{Llama 2} & \begin{tabular}[c]{@{}l@{}}temperature: 0.01\\ max\_new\_tokens: 200\\ top\_p: 0.95\\ top\_k: 250\\ length\_penalty: 1\\ presence\_penalty: 0\end{tabular} \\ \hline
\textbf{Mistral 7B} & \begin{tabular}[c]{@{}l@{}}temperature: 0.01\\ max\_new\_tokens: 128\\ top\_p: 0.95\\ top\_k: 50\\ length\_penalty: 1\\ presence\_penalty: 0\end{tabular} \\ \hline
\textbf{Gemma} & \begin{tabular}[c]{@{}l@{}}temperature: 0.01\\ max\_new\_tokens: 200\end{tabular} \\ \hline
\textbf{Phi 1.5} & \begin{tabular}[c]{@{}l@{}}temperature: 0.01\\ max\_length: 500\end{tabular} \\ \hline
\textbf{Phi 2} & \begin{tabular}[c]{@{}l@{}}temperature: 0.01\\ max\_length: 500\end{tabular} \\ \hline
\textbf{Phi 3} & \begin{tabular}[c]{@{}l@{}}temperature: 0.01\\ max\_length: 500\end{tabular} \\ \hline
\textbf{GPT-J} & \begin{tabular}[c]{@{}l@{}}temperature: 0.01\\ max\_new\_tokens: 200\end{tabular} \\ \hline
\textbf{Dolly} & \begin{tabular}[c]{@{}l@{}}temperature: 0.01\\ max\_new\_tokens: 200\end{tabular} \\ \hline
\end{tabular}%
}
\label{tab:LLM_hyperparameters}
\end{table}

\newpage 

\section{Prompt Example}
\label{app:full_prompt}

\begin{longtable}{|p{0.5\linewidth}|}
    \caption{Zero-shot prompt for next location prediction used for all the LLMs}
    \label{tab:prompt} \\
    \hline
    \textbf{Prompt Zero-Shot} \\
    \hline
    Your task is to predict \texttt{<next\_place\_id>} in \texttt{<target\_stay>}, a location with an unknown ID, while temporal data is available. \newline
    \newline
    Predict \texttt{<next\_place\_id>} by considering: \newline
    1. The user's activity trends gleaned from \texttt{<historical\_stays>} and the current activities from \texttt{<context\_stays>}. \newline
    2. Temporal details (start\_time and day\_of\_week) of the target stay, crucial for understanding activity variations. \newline
    \newline
    Present your answer in a JSON object with: \newline
    \texttt{"prediction"} (IDs of the five most probable places, ranked by probability) and \texttt{"reason"} (a concise justification for your prediction). \newline
    \newline
    The data: \newline
    \texttt{<historical\_stays>}: \newline
    \{['6 PM', 'Sunday', '42ec1480f964a5209e261fe3'], \newline
      ['1 PM', 'Monday', '4d48864b551db1f722e9a330'], \newline
      ['1 PM', 'Wednesday', '4d48864b551db1f722e9a330'], \newline
      ['4 PM', 'Wednesday', '4cdf0650ffcf37041bd71682'], \newline
      ['5 PM', 'Wednesday', '4b705cd2f964a52052152de3'], \newline
      ['5 PM', 'Wednesday', '4d531225f1b4a09039dee672'], \newline
      ['5 PM', 'Wednesday', '4ada106af964a520ed1d21e3'], \newline
      ['6 PM', 'Wednesday', '4c176109f693b7139f0f5f54'], \newline
      ['2 AM', 'Friday', '4c498ca56f1420a1c019c155'], \newline
      ['1 AM', 'Saturday', '4b7c698af964a520bc912fe3'], \newline
      ['2 PM', 'Saturday', '41102700f964a520d60b1fe3'], \newline
      ['8 PM', 'Saturday', '4be5dfe9910020a13ea2d314'], \newline
      ['8 PM', 'Saturday', '4a43c0aef964a520c6a61fe3'], \newline
      ['4 PM', 'Sunday', '41102700f964a520d60b1fe3'], \newline
      ['5 PM', 'Monday', '4ba7ef65f964a520fabf39e3']\} \newline
    \texttt{<context\_stays>}: \newline
    \{['5 PM', 'Wednesday', '4531059cf964a520683b1fe3'], \newline
      ['6 PM', 'Wednesday', '50a430bfe4b0f010d751e3a5'], \newline
      ['7 PM', 'Wednesday', '4a871b30f964a5209a0220e3'], \newline
      ['3 PM', 'Thursday', '4b8f50d9f964a520365133e3'], \newline
      ['11 PM', 'Thursday', '4e7bb781887742fc66380d56'], \newline
      ['7 PM', 'Friday', '4b7ecc2cf964a520420030e3']\} \newline
    \texttt{<target\_stay>}: \newline
    \{<next\_place\_id'>\} \\
    \hline
\end{longtable}

\newpage
\begin{longtable}{|p{0.95\linewidth}|}
    \caption{One-shot prompt for next location prediction used for all the LLMs}
    \label{tab: One-Shot prompt} \\
    \hline
    \textbf{Prompt One-Shot} \\
    \hline
Original Prompt: \\
    Your task is to predict \texttt{<next\_place\_id>} in \texttt{<target\_stay>}, a location with an unknown ID, while temporal data is available. \newline
    \newline
    Predict \texttt{<next\_place\_id>} by considering: \newline
    1. The user's activity trends gleaned from \texttt{<historical\_stays>} and the current activities from \texttt{<context\_stays>}. \newline
    2. Temporal details (start\_time and day\_of\_week) of the target stay, crucial for understanding activity variations. \newline
    \newline
    Present your answer in a JSON object with: \newline
    \texttt{"prediction"} (IDs of the five most probable places, ranked by probability) and \texttt{"reason"} (a concise justification for your prediction). \newline
    \newline
    The data: \newline
    \texttt{<historical\_stays>}: \newline
    \{['6 PM', 'Sunday', '42ec1480f964a5209e261fe3'], \newline
      ['1 PM', 'Monday', '4d48864b551db1f722e9a330'], \newline
      ['1 PM', 'Wednesday', '4d48864b551db1f722e9a330'], \newline
      ['4 PM', 'Wednesday', '4cdf0650ffcf37041bd71682'], \newline
      ['5 PM', 'Wednesday', '4b705cd2f964a52052152de3'], \newline
      ['5 PM', 'Wednesday', '4d531225f1b4a09039dee672'], \newline
      ['5 PM', 'Wednesday', '4ada106af964a520ed1d21e3'], \newline
      ['6 PM', 'Wednesday', '4c176109f693b7139f0f5f54'], \newline
      ['2 AM', 'Friday', '4c498ca56f1420a1c019c155'], \newline
      ['1 AM', 'Saturday', '4b7c698af964a520bc912fe3'], \newline
      ['2 PM', 'Saturday', '41102700f964a520d60b1fe3'], \newline
      ['8 PM', 'Saturday', '4be5dfe9910020a13ea2d314'], \newline
      ['8 PM', 'Saturday', '4a43c0aef964a520c6a61fe3'], \newline
      ['4 PM', 'Sunday', '41102700f964a520d60b1fe3'], \newline
      ['5 PM', 'Monday', '4ba7ef65f964a520fabf39e3']\} \newline
    \texttt{<context\_stays>}: \newline
    \{['5 PM', 'Wednesday', '4531059cf964a520683b1fe3'], \newline
      ['6 PM', 'Wednesday', '50a430bfe4b0f010d751e3a5'], \newline
      ['7 PM', 'Wednesday', '4a871b30f964a5209a0220e3'], \newline
      ['3 PM', 'Thursday', '4b8f50d9f964a520365133e3'], \newline
      ['11 PM', 'Thursday', '4e7bb781887742fc66380d56'], \newline
      ['7 PM', 'Friday', '4b7ecc2cf964a520420030e3']\} \newline
    \texttt{<target\_stay>}: \newline
    \{<next\_place\_id'>\} \\
    prediction: \newline
    \{['4b80bafef964a520ee8830e3',\newline '4b558306f964a5201be627e3',\newline '4b7e3467f964a520fde52fe3',\newline '4c182e2c6a21c9b6f2bbc897',\newline '4dff7dd51495f702193690bf']\}, \newline
    reason: \newline
    'User has repeatedly visited places in the evening and at night, especially on weekdays. The most recent context stays are also at nighttime.' \newline
    \newline
    Original prompt: \newline
    Your task is to predict \texttt{<next\_place\_id>} in \texttt{<target\_stay>}, a location with an unknown ID, while temporal data is available. \newline
    \newline
    Predict \texttt{<next\_place\_id>} by considering: \newline
    1. The user's activity trends gleaned from \texttt{<historical\_stays>} and the current activities from \texttt{<context\_stays>}. \newline
    2. Temporal details (start\_time and day\_of\_week) of the target stay, crucial for understanding activity variations. \newline
    \newline
    Present your answer in a JSON object with: \newline
    \texttt{"prediction"} (IDs of the five most probable places, ranked by probability) and \texttt{"reason"} (a concise justification for your prediction).
    \\
The data: \newline
\texttt{<historical\_stays>}: \newline
\{['6 PM', 'Sunday', '42ec1480f964a5209e261fe3'], \newline
  ['1 PM', 'Monday', '4d48864b551db1f722e9a330'], \newline
  ['1 PM', 'Wednesday', '4d48864b551db1f722e9a330'], \newline
  ['4 PM', 'Wednesday', '4cdf0650ffcf37041bd71682'], \newline
  ['5 PM', 'Wednesday', '4b705cd2f964a52052152de3'], \newline
  ['5 PM', 'Wednesday', '4d531225f1b4a09039dee672'], \newline
  ['5 PM', 'Wednesday', '4ada106af964a520ed1d21e3'], \newline
  ['6 PM', 'Wednesday', '4c176109f693b7139f0f5f54'], \newline
  ['2 AM', 'Friday', '4c498ca56f1420a1c019c155'], \newline
  ['1 AM', 'Saturday', '4b7c698af964a520bc912fe3'], \newline
  ['2 PM', 'Saturday', '41102700f964a520d60b1fe3'], \newline
  ['8 PM', 'Saturday', '4be5dfe9910020a13ea2d314'], \newline
  ['8 PM', 'Saturday', '4a43c0aef964a520c6a61fe3'], \newline
  ['4 PM', 'Sunday', '41102700f964a520d60b1fe3'], \newline
  ['5 PM', 'Monday', '4ba7ef65f964a520fabf39e3']\} \newline
\texttt{<context\_stays>}: \newline
\{['3 AM', 'Friday', '44166f04f964a52003311fe3'], \newline
  ['11 PM', 'Friday', '4b55f1eef964a52002f827e3'], \newline
  ['7 PM', 'Saturday', '4c537ec4fd2ea5936ffe8328']\} \newline
\texttt{<target\_stay>}: \newline
\{<next\_place\_id'>\} \\
    \hline
\end{longtable}

\newpage 
\begin{longtable}{|p{0.95\linewidth}|}
    \caption{Few-Shot prompt for next location prediction used for all the LLMs}
    \label{tab:Few-Shot prompt} \\
    \hline
    \textbf{Prompt Few-Shot} \\
    \hline
    Your task is to predict \texttt{<next\_place\_id>} in \texttt{<target\_stay>}, a location with an unknown ID, while temporal data is available. Predict \texttt{<next\_place\_id>} by considering: \\
    1. The user's activity trends gleaned from \texttt{<historical\_stays>} and the current activities from \texttt{<context\_stays>}. \\
    2. Temporal details (start\_time and day\_of\_week) of the target stay, crucial for understanding activity variations. \\
    Present your answer in a JSON object with: \texttt{"prediction"} (IDs of the five most probable places, ranked by probability) and \texttt{"reason"} (a concise justification for your prediction). \\
    Consider the following examples to understand the user's patterns: \\
    Example 1: \\
    \textbf{historical\_stays}: \{['10:49 PM', 'Sunday', '4b80bafef964a520ee8830e3'], ['03:16 AM', 'Monday', '4c182e2c6a21c9b6f2bbc897'], ['08:57 AM', 'Monday', '4b8ef710f964a5209c4133e3'], ['02:46 PM', 'Monday', '4b80bafef964a520ee8830e3'], ['03:13 PM', 'Wednesday', '4b80bafef964a520ee8830e3'], ['02:10 AM', 'Thursday', '4bb365f54019a593e6d937b8'], ['11:29 PM', 'Sunday', '4b80bafef964a520ee8830e3'], ['12:32 AM', 'Monday', '4bb365f54019a593e6d937b8'], ['12:59 AM', 'Monday', '4b7e3467f964a520fde52fe3'], ['02:22 PM', 'Thursday', '4b5d54a2f964a5200e5a29e3'], ['01:40 PM', 'Friday', '4cdbe66e22bd721e4302f847'], ['02:03 PM', 'Saturday', '4e13debbe4cd473c968b5afc'], ['12:17 AM', 'Friday', '4bf5d425004ed13aa27541a0'], ['11:04 AM', 'Friday', '4c2c8fd677cfe21e2029b6f1'], ['08:58 AM', 'Thursday', '4b7e3467f964a520fde52fe3'], ['07:26 AM', 'Friday', '4dff7dd51495f702193690bf'], ['07:32 AM', 'Friday', '4bb365f54019a593e6d937b8'], ['07:33 AM', 'Friday', '4b7e3467f964a520fde52fe3'], ['12:19 AM', 'Tuesday', '4bf5d425004ed13aa27541a0']\} \\
    \textbf{context\_stays}: \{['12:09 AM', 'Wednesday', '4dff7dd51495f702193690bf'], ['10:44 AM', 'Wednesday', '4b8da54ef964a5202e0633e3'], ['11:04 AM', 'Wednesday', '4dff7dd51495f702193690bf'], ['11:12 AM', 'Wednesday', '4b7e3467f964a520fde52fe3'], ['02:24 PM', 'Wednesday', '4b22e836f964a520185024e3']\} \\
    \textbf{target\_stay}: \{'prediction': ['4b80bafef964a520ee8830e3', '4b558306f964a5201be627e3', '4b7e3467f964a520fde52fe3', '4c182e2c6a21c9b6f2bbc897', '4dff7dd51495f702193690bf']\} \\
    Example 2: \\
    \textbf{historical\_stays}: \{['03:47 AM', 'Monday', '4b7e3467f964a520fde52fe3'], ['10:45 PM', 'Tuesday', '4b80bafef964a520ee8830e3'], ['11:39 PM', 'Tuesday', '4dff7dd51495f702193690bf'], ['11:45 PM', 'Tuesday', '4bb365f54019a593e6d937b8'], ['03:01 PM', 'Thursday', '4b22e836f964a520185024e3']\} \\
    \textbf{context\_stays}: \{['12:09 AM', 'Wednesday', '4dff7dd51495f702193690bf'], ['10:44 AM', 'Wednesday', '4b8da54ef964a5202e0633e3'], ['11:04 AM', 'Wednesday', '4dff7dd51495f702193690bf'], ['11:12 AM', 'Wednesday', '4b7e3467f964a520fde52fe3'], ['02:24 PM', 'Wednesday', '4b22e836f964a520185024e3']\} \\
    \textbf{target\_stay}: \{'prediction': ['4b6ba709f964a52059142ce3', '4c182e2c6a21c9b6f2bbc897', '4b8ef710f964a5209c4133e3', '4bea89ca415e20a1af16e5bb', '4dff7dd51495f702193690bf']\} \\
    Consider the user's activity patterns, historical stays, and recent context stays when making your prediction. Consider also the temporal information of the target stay. Provide the 5 most likely places (\texttt{place\_id}) ranked in descending order of probability. Ensure that your output does not include line breaks. \\
    The data are as follows: \\
    \texttt{<historical\_stays>}: \{['9 AM', 'Tuesday', '50e8ed15e4b07ff4593e1bf4'], ['10 AM', 'Wednesday', '4dd2d4bc81dcfc9dcdc2124f'], ['10 AM', 'Wednesday', '4df8089a1f6e448ec28e6fe0'], ['12 PM', 'Wednesday', '4d22782eb69c6dcbf7be8295'], ['12 PM', 'Wednesday', '4da61b1ea86ecc5168c8828c'], ['2 PM', 'Wednesday', '4bf804ee8d30d13a1feaff17'], ['8 AM', 'Thursday', '4bae676cf964a520c6ad3be3'], ['9 AM', 'Thursday', '4eba3331722edc0eaf1762bb'], ['4 AM', 'Friday', '4bd1af9b41b9ef3bca54fce5'], ['5 AM', 'Friday', '4b5d272ef964a520c15429e3'], ['7 AM', 'Monday', '4c7f04be7fd4ef3b108394a1'], ['0 AM', 'Tuesday', '4d9fa1d483f0b1f7ea8c9ec7'], ['1 AM', 'Tuesday', '4c4b52ce9e6dbe9acbd4e10b'], ['8 AM', 'Tuesday', '4c7f04be7fd4ef3b108394a1'], ['12 PM', 'Tuesday', '4b06d110f964a520c3f022e3']\} \\
    \texttt{<context\_stays>}: \{['7 PM', 'Saturday', '510c9f9de4b0ec5cc005fc87'], ['8 PM', 'Saturday', '4bae0e67f964a520487f3be3'], ['8 PM', 'Saturday', '4be9e20e6295c9b61acb8508'], ['9 PM', 'Saturday', '4d1fdddf5c4ca1cdb4a49c3d'], ['0 AM', 'Sunday', '4cc744c0a32bb1f75758a7a8'], ['0 AM', 'Sunday', '4abd30e1f964a520ed8820e3'], ['2 AM', 'Sunday', '4b60b7e5f964a520b7f529e3'], ['4 AM', 'Sunday', '4bea19226295c9b6b61e8608'], ['5 AM', 'Sunday', '4cc184d3f82ebfb75c417091'], ['6 PM', 'Monday', '4eba3331722edc0eaf1762bb'], ['2 AM', 'Tuesday', '4c65b90018b676b084a0ad0e'], ['4 AM', 'Tuesday', '501b374ce4b06c31c5001715'], ['6 AM', 'Tuesday', '4cfeed7c202e224bc2a7ab2a'], ['8 AM', 'Tuesday', '4c7f04be7fd4ef3b108394a1'], ['9 AM', 'Tuesday', '4b0737c3f964a520e4f922e3'], ['10 AM', 'Tuesday', '4dd3fa5e6365c27b0dbeec7d'], ['11 AM', 'Tuesday', '4a257482f964a520557e1fe3'], ['12 PM', 'Tuesday', '4cd78a30a42b236ad848fb08']\} \\
    \texttt{<target\_stay>}: \{<next\_place\_id'>\} \\
    \hline
\end{longtable}

\newpage 
\subsection{Output Examples}

\begin{longtable}{|p{0.95\linewidth}|}
    \caption{GPT-4o output example}
    \label{tab:GPT-4o_output} \\
    \hline
    \textbf{GPT-4o} \\
    \hline
    \lstset{
        basicstyle=\ttfamily\footnotesize,
        breaklines=true
    }
    \begin{lstlisting}
{
        "prediction": [
            "4a5dd98ef964a520c2bd1fe3",
            "3fd66200f964a520def11ee3",
            "4a3d735cf964a5204ba21fe3",
            "4d48864b551db1f722e9a330",
            "42ec1480f964a5209e261fe3"
        ],
        "reason": "The user has frequently visited the places '4d48864b551db1f722e9a330', \\ '4a5dd98ef964a520c2bd1fe3', '3fd66200f964a520def11ee3', '42ec1480f964a5209e261fe3', and \\ '4a3d735cf964a5204ba21fe3'. In addition, context locations provide recent activities at places \\ '4a5dd98ef964a520c2bd1fe3', '3fd66200f964a520def11ee3', and '4a3d735cf964a5204ba21fe3'. \\ It is likely that the user will visit one of these places again."
    }
    \end{lstlisting} \\
    \hline
\end{longtable}

\begin{longtable}{|p{0.95\linewidth}|}
    \caption{GPT-4 output example}
    \label{tab:GPT-4_output} \\
    \hline
    \textbf{GPT-4} \\
    \hline
    \lstset{
        basicstyle=\ttfamily\footnotesize,
        breaklines=true
    }
    \begin{lstlisting}
{
        "prediction": [
            "4b7ecc2cf964a520420030e3",
            "41102700f964a520d60b1fe3",
            "4d531225f1b4a09039dee672",
            "4b705cd2f964a52052152de3",
            "4c176109f693b7139f0f5f54"
        ],
        "reason": "Based on the historical stays, the user has frequently visited the places with place IDs 4b7ecc2cf964a520420030e3, 41102700f964a520d60b1fe3, 4d531225f1b4a09039dee672, 4b705cd2f964a52052152de3, and 4c176109f693b7139f0f5f54. These places have been visited during different times of the day and on different days of the week. Considering the current context stays and the temporal information of the target stay, it is likely that the user will visit one of these places as his/her next location."
    }
    \end{lstlisting} \\
    \hline
\end{longtable}

\begin{longtable}{|p{0.95\linewidth}|}
    \caption{GPT-3.5 output example}
    \label{tab:GPT-3.5_output} \\
    \hline
    \textbf{GPT-3.5 Output} \\
    \hline
    \lstset{
        basicstyle=\ttfamily\footnotesize,
        breaklines=true
    }
    \begin{lstlisting}
{
        "prediction": [
            "4b7ecc2cf964a520420030e3",
            "41102700f964a520d60b1fe3",
            "4d531225f1b4a09039dee672",
            "4b705cd2f964a52052152de3",
            "4c176109f693b7139f0f5f54"
        ],
        "reason": "Based on the historical stays, the user has frequently visited the places with place IDs 4b7ecc2cf964a520420030e3, 41102700f964a520d60b1fe3, 4d531225f1b4a09039dee672, 4b705cd2f964a52052152de3, and 4c176109f693b7139f0f5f54. These places have been visited during different times of the day and on different days of the week. Considering the current context stays, it is likely that the user will visit one of these places as his/her next location."
    }
    \end{lstlisting} \\
    \hline
\end{longtable}

\begin{longtable}{|p{0.95\linewidth}|}
    \caption{Llama 3.1 8B output example}
    \label{tabLlama 3.1 8B_output} \\
    \hline
    \textbf{Llama 3.1 8B} \\
    \hline
    \lstset{
        basicstyle=\ttfamily\footnotesize,
        breaklines=true
    }
    \begin{lstlisting}
["4b9ae820f964a520c7e135e3" ,"4e89a96d4690c079de1c85f8", "4c3b49c54565e21eb775566a", "4bb1fa0bf964a52010af3ce3", "4e89a96d4690c079de1c85f8"]. Considering historical visits, user will likely visit recurrent places."
    \end{lstlisting} \\
    \hline
\end{longtable}

\begin{longtable}{|p{0.95\linewidth}|}
    \caption{Llama 3 70B Instruct  output example}
    \label{tab:Llama 3 70B Instruct  _output} \\
    \hline
    \textbf{Llama 3 70B Instruct } \\
    \hline
    \lstset{
        basicstyle=\ttfamily\footnotesize,
        breaklines=true
    }
    \begin{lstlisting}
{"prediction": 
["4be5dfe9910020a13ea2d314",
"4c176109f693b7139f0f5f54",
"4d531225f1b4a09039dee672", 
"4c176109f693b7139f0f5f54", 
"4b705cd2f964a52052152de3"],
"reason": "Based on the user's historical stays, it is likely that they will visit a place they have been to before. The top five predicted places are all locations that frequently visited, which suggests a pattern of regular activity at these times"
    \end{lstlisting} \\
    \hline
\end{longtable}

\begin{longtable}{|p{0.95\linewidth}|}
    \caption{Llama 3 8B Instruct  output example}
    \label{tab:Llama 3 70B Instruct  _output} \\
    \hline
    \textbf{Llama 3 8B Instruct } \\
    \hline
    \lstset{
        basicstyle=\ttfamily\footnotesize,
        breaklines=true
    }
    \begin{lstlisting}
"prediction": {"3fd66200f964a520cce51ee3", "4ce08a08ffcf370469b12082", "4ad507bff964a520660121e3", "4c881d60a0ffb60ccda81fc5"},"reason": "Based on the user's historical stays and contextual information, it is likely that the user will visit a place similar to previous destinations.
    \end{lstlisting} \\
    \hline
\end{longtable}

\begin{longtable}{|p{0.95\linewidth}|}
    \caption{Llama 3 70B output example}
    \label{tabLlama 3.1 8B_output} \\
    \hline
    \textbf{Llama 3 70B} \\
    \hline
    \lstset{
        basicstyle=\ttfamily\footnotesize,
        breaklines=true
    }
    \begin{lstlisting}
["4b4b6e24f964a5204f9b26e3","4a8ddf35f964a520341120e3","4cdac798d6656a315f36fc3e","4cdac798d6656a315f36fc3e","4cdac798d6656a315f36fc3e"] Contextural visits are recurrent and historical visits indicate the user will likely visits similar locations.
    \end{lstlisting} \\
    \hline
\end{longtable}

\begin{longtable}{|p{0.95\linewidth}|}
    \caption{Llama 3  8B output example}
    \label{tabLlama 3.1 8B_output} \\
    \hline
    \textbf{Llama 3 8B} \\
    \hline
    \lstset{
        basicstyle=\ttfamily\footnotesize,
        breaklines=true
    }
    \begin{lstlisting}
["4dc326182271f27050f39597" ,"4bb1fa0bf964a52010af3ce3", "4bfaa7ed5ec320a1d2f28bd3", "4f7897ace4b0421083e165f8", "4ae0d39bf964a520e68221e3"]. User will likely visit places with
high repeat visits "

    \end{lstlisting} \\
    \hline
\end{longtable}

\begin{longtable}{|p{0.95\linewidth}|}
    \caption{Llama 2 Chat 70B output example}
    \label{tab:Llama 2 Chat 70B_output} \\
    \hline
    \textbf{Llama 2 Chat 70B Output} \\
    \hline
    \lstset{
        basicstyle=\ttfamily\footnotesize,
        breaklines=true
    }
    \begin{lstlisting}
{"prediction": 
["4be5dfe9910020a13ea2d314",
"4ada106af964a520ed1d21e3",
"4cdf0650ffcf37041bd71682", 
"4c176109f693b7139f0f5f54", 
"4ba7ef65f964a520fabf39e3"],
"reason": "Based on the user's historical stays, it is likely that they will visit a place they have been to before, especially during the same time of day and day of the week. The top five predicted places are all locations that the user has previously visited on Wednesdays, which suggests a pattern of regular activity at these times"
    }
    \end{lstlisting} \\
    \hline
\end{longtable}

\newpage
\begin{longtable}{|p{0.95\linewidth}|}
    \caption{Llama 2 Chat 13B output example}
    \label{tab:Llama 2 Chat 13B_output} \\
    \hline
    \textbf{Llama 2 Chat 13B Output} \\
    \hline
    \lstset{
        basicstyle=\ttfamily\footnotesize,
        breaklines=true
    }
    \begin{lstlisting}
{" Sure, I'd be happy to help! Based on the provided data, here is my prediction for the next location of the user:"prediction": ["4c176109f693b7139f0f5f54","4d48864b551db1f722e9a330","4ada106af964a520ed1d21e3","4ba7ef65f964a520fabf39e3","4be5dfe9910020a13ea2d314"],"reason": "Based on the user's historical stays, it is likely that they will visit places with high repeat visits and proximity to their recent context stays. The target"
    }
    \end{lstlisting} \\
    \hline
\end{longtable}

\begin{longtable}{|p{0.95\linewidth}|}
    \caption{Llama 2 Chat 7B output example}
    \label{tab:Llama 2 Chat 7B_output} \\
    \hline
    \textbf{Llama 2 Chat 7B Output} \\
    \hline
    \lstset{
        basicstyle=\ttfamily\footnotesize,
        breaklines=true
    }
    \begin{lstlisting}
{"Based on the provided data, I have predicted the next location of the user as follows:"prediction": ["4af5a46af964a520b5fa21e3", "4b0a9290f964a520352523e3", "4ad507bff964a520660121e3", "4c881d60a0ffb60ccda81fc5"],"reason": "Based on the user's historical stays and contextual information, it is likely that the user will visit a place that is similar to their previous destinations.}
    \end{lstlisting} \\
    \hline
\end{longtable}

\begin{longtable}{|p{0.95\linewidth}|}
    \caption{Llama 2 70B output example}
    \label{tab:Llama 2  70B_output} \\
    \hline
    \textbf{Llama 2 70B Output} \\
    \hline
    \lstset{
        basicstyle=\ttfamily\footnotesize,
        breaklines=true
    }
    \begin{lstlisting}    
["4c498ca56f1420a1c019c155","4d531225f1b4a09039dee672","4b7c698af964a520bc912fe3","4d531225f1b4a09039dee672","4b7c698af964a520bc912fe3"] Based on the user's historical stays, it is likely that they will visit places with high repeat visits"
    \end{lstlisting} \\
    \hline
\end{longtable}

\begin{longtable}{|p{0.95\linewidth}|}
    \caption{Llama 2 13B output example}
    \label{tab:Llama 2 13B_output} \\
    \hline
    \textbf{Llama 2 13B Output} \\
    \hline
    \lstset{
        basicstyle=\ttfamily\footnotesize,
        breaklines=true
    }
    \begin{lstlisting}       
["41102700f964a520d60b1fe3","41102700f964a520d60b1fe3","4ada106af964a520ed1d21e3","4d531225f1b4a09039dee672","4d531225f1b4a09039dee672"] User will likely visit places with high repeat visits"
    \end{lstlisting} \\
    \hline
\end{longtable}

\newpage
\begin{longtable}{|p{0.95\linewidth}|}
    \caption{Llama 2 7B output example}
    \label{tab:Llama 2  7B_output} \\
    \hline
    \textbf{Llama 2 7B Output} \\
    \hline
    \lstset{
        basicstyle=\ttfamily\footnotesize,
        breaklines=true
    }
    \begin{lstlisting}
["4b7ecc2cf964a520420030e3","4b7ecc2cf964a520420030e3","4b7ecc2cf964a520420030e3","4b7ecc2cf964a520420030e3","4b7ecc2cf964a520420030e3"] User will likely visit places with high repeat visits
    \end{lstlisting} \\
    \hline
\end{longtable}

\begin{longtable}{|p{0.95\linewidth}|}
    \caption{Mistral 7B output example}
    \label{tab:Mistral 7B_output} \\
    \hline
    \textbf{Mistral 7B Output} \\
    \hline
    \lstset{
        basicstyle=\ttfamily\footnotesize,
        breaklines=true
    }
    \begin{lstlisting}
        The user frequently visits ['4be5dfe9910020a13ea2d314', '4a5dd98ef964a520c2bd1fe3', '4a43c0aef964a520c6a61fe3', '4a3d735cf964a5204ba21fe3', '4a3d735cf964a5204ba21fe3']. The user will likely visit one of these places again.
    \end{lstlisting} \\
    \hline
\end{longtable}

\subsection{Failed Models}

\begin{longtable}{|p{0.95\linewidth}|}
    \caption{Dolly 3B output example}
    \label{tab:Dolly_3B_output} \\
    \hline
    \textbf{Dolly 3B Output} \\
    \hline
    \lstset{
        basicstyle=\ttfamily\footnotesize,
        breaklines=true
    }
    \begin{lstlisting}
{"prediction": "Your response will contain the following keys:   "prediction" and "reason"  The JSON will contain the following key/value pair:   "prediction": ["14687", "Night Life"]   "reason": "According to our stay history, the probability that the place 14687 is the next location is high."  The output will be as follows:   {"prediction": ["14687", "Night Life"]}  Note that the location 14687 is merely an example. In reality, it might contain more than 10 possible locations where this user may go next. Please see the following example for more details on how to choose the most probable place.      {       "prediction": ["14687", "Night Life"],       "reason": "According to our stay history, the probability that the place 14687 is the next location is high."   }  That's it!  PS. This answer was submitted in about 1 min 36 secs, using JSON formatted data provided.", "reason": null}
    \end{lstlisting} \\
    \hline
\end{longtable}

\newpage
\begin{longtable}{|p{0.95\linewidth}|}
    \caption{Dolly 7B output example}
    \label{tab:Dolly_7B_output} \\
    \hline
    \textbf{Dolly 7B Output} \\
    \hline
    \lstset{
        basicstyle=\ttfamily\footnotesize,
        breaklines=true
    }
    \begin{lstlisting}
{    "prediction": [        "04:54 AM",        "10:49 PM",        "11:52 PM",        "12:09 AM",        "02:13 AM",        "02:13 AM",        "07:26 AM",        "08:33 PM",        "03:39 AM",        "06:51 PM",        "12:15 AM",        "03:21 AM",        "10:15 PM",        "01:40 AM",        "02:33 AM",        "07:35 AM",        "05:31 PM",        "09:08 PM",        "11:24 PM",        "11:55 PM",        "11:55 PM",        "01:11 AM",        "06:46 AM",        "07:02 PM",        "12:30 AM",        "06:02 AM",        "05:45 PM",        "07:02 AM",        "07:12 PM",        "06:42 AM",        "08:17 AM", "reason": null}
    \end{lstlisting} \\
    \hline
\end{longtable}

\begin{longtable}{|p{0.95\linewidth}|}
    \caption{Dolly 12B output example}
    \label{tab:Dolly_12B_output} \\
    \hline
    \textbf{Dolly 12B Output} \\
    \hline
    \lstset{
        basicstyle=\ttfamily\footnotesize,
        breaklines=true
    }
    \begin{lstlisting}
" prediction ": "10:29 PM " , " reason ": "
most likely " } , { "
prediction ": "12:43 AM " , " reason ": "
most likely " } , { "
prediction ": "12:44 AM " , " reason ": "
most likely " } ]  The most probable
next location is "12:43 AM " for this user , since "
historical_stays " show this as their most recent activity and
there are three historical stay with an overlap of 11:06 PM to
12:43 AM . This can also be explained with the temporal data in
the " target_stay " as this is the only stay with an overlap
of 11:06 PM to 12:43 AM .  This supports the users most
recent activity of "11:06 PM " for the following reasons :
1 . This is the most recent activity of a " Tuesday " and
this is the most recent activity of any day .2 . This is also
the most recent activity of a " Wednesday " and this is the
most recent activity of any day .3 . This is also the most " , "
reason ": null

    \end{lstlisting} \\
    \hline
\end{longtable}

\begin{longtable}{|p{0.95\linewidth}|}
    \caption{GPT-J output example}
    \label{tab:GPT-J_output} \\
    \hline
    \textbf{GPT-J Output} \\
    \hline
    \lstset{
        basicstyle=\ttfamily\footnotesize,
        breaklines=true
    }
    \begin{lstlisting}
The following code snippet shows this data parsing.```python# Importing required librariesimport jsonimport timeimport randomimport numpy as npimport pandas as pd# Read JSON datawith open('data.json', 'r') as json_input:    json_data = json.load(json_input)# Transform data into Pandas DataFrame formatdata = pd.DataFrame.from_dict(json_data['historical_stays'])# The following two lines are for testing only#data = pd.DataFrame([[] for k in range(data.shape[0])])#data['reason'] = {k:[] for k in range(data.shape[0])}# Defining variablespredicted_idx = random.sample(range(1, 11), 10)data = data.iloc[predicted_idx,:]# Extract unique place IDs from each stayplace_ids = [data['place_id'].apply(lambda x:", "reason": null}
    \end{lstlisting} \\
    \hline
\end{longtable}

\begin{longtable}{|p{0.95\linewidth}|}
    \caption{Phi 1.5 output example}
    \label{tab:Phi 1.5_output} \\
    \hline
    \textbf{Phi 1.5 Output} \\
    \hline
    \lstset{
        basicstyle=\ttfamily\footnotesize,
        breaklines=true
    }
    \begin{lstlisting}
Your task is to predict a user's next location based on his/her activity pattern.                          You will be provided with a JSON containing <historical_stays> which contain this user's historical stays, then <context_stays> which provide contextual information
    \end{lstlisting} \\
    \hline
\end{longtable}

\begin{longtable}{|p{0.95\linewidth}|}
    \caption{Phi 2 output example}
    \label{tab:Phi 2 _output} \\
    \hline
    \textbf{Phi 2 Output} \\
    \hline
    \lstset{
        basicstyle=\ttfamily\footnotesize,
        breaklines=true
    }
    \begin{lstlisting}
{[place]}\\
    \end{lstlisting} \\
    \hline
\end{longtable}

\begin{longtable}{|p{0.95\linewidth}|}
    \caption{Phi 3 output example}
    \label{tab:Phi 3_output} \\
    \hline
    \textbf{Phi 3 Output} \\
    \hline
    \lstset{
        basicstyle=\ttfamily\footnotesize,
        breaklines=true
    }
    \begin{lstlisting}
prediction \\
    \end{lstlisting} \\
    \hline
\end{longtable}

\begin{longtable}{|p{0.95\linewidth}|}
    \caption{Dolly 2B output example}
    \label{tab:Dolly 2B_output} \\
    \hline
    \textbf{Dolly 2B Output} \\
    \hline
    \lstset{
        basicstyle=\ttfamily\footnotesize,
        breaklines=true
    }
    \begin{lstlisting}


**Output:**

```json
{"prediction": 2, "reason": "People tend to visit the same places at similar times throughout the week, so the next most probable place would be somewhere within that vicinity."}
```
\end{lstlisting} \\
    \hline
\end{longtable}

\newpage
\subsection{Hallucinated output example}

\begin{longtable}{|p{0.95\textwidth}|}
    \caption{This table illustrates an example of hallucinated output generated by GPT-3.5. The output appears coherent and includes location IDs similar to the Foursquare dataset. However, these IDs do not match the input data, the original dataset, or any valid Foursquare POIs.}
    \label{tab:HallucinatedOutput} \\
    \hline
    \textbf{Hallucinated Output} \\
    \hline
    \lstset{
        basicstyle=\ttfamily\footnotesize,  
        breaklines=true,
        columns=fullflexible,
        keepspaces=true,
        showstringspaces=false,
        breakatwhitespace=true,
        frame=none
    }
    \begin{lstlisting}
"input": " Your task is to predict <next_place_id> in <target_stay>, a location with an unknown ID,
while temporal data is available.
Predict <next_place_id> by considering:
1. The user's activity trends gleaned from <historical_stays> and the current activities
from <context_stays>.
2. Temporal details (start_time and day_of_week) of the target stay, crucial for understanding
activity variations.
Present your answer in a JSON object with:
'prediction' (IDs of the five most probable places, ranked by probability) and 'reason' (a
concise justification for your prediction).
the data:

  <historical_stays>: 
  [['0 AM', 'Monday', '4b00383ff964a5209f3b22e3'], ['1 AM', 'Thursday', '3fd66200f964a5203de71ee3'], ['1 AM', 'Saturday', '3fd66200f964a52005eb1ee3'], ['1 AM', 'Sunday', '44ddb091f964a520bf361fe3'], ['3 AM', 'Sunday', '4aa48305f964a520004720e3'], ['0 AM', 'Monday', '45699170f964a520e73d1fe3'], ['5 PM', 'Thursday', '50036b8ee4b04c8e75c1945f'], ['5 PM', 'Saturday', '4f900989e4b0324e976f087b'], ['8 PM', 'Saturday', '3fd66200f964a520cce51ee3'], ['0 AM', 'Monday', '4a5e4a0df964a52050be1fe3'], ['2 AM', 'Tuesday', '43a26468f964a520322c1fe3'], ['6 PM', 'Friday', '4ce08a08ffcf370469b12082'], ['11 PM', 'Friday', '46c77917f964a5201a4a1fe3'], ['6 PM', 'Saturday', '4a47de2df964a52036aa1fe3'], ['8 PM', 'Saturday', '4b9ee138f964a5200d0837e3']]
  <context_stays>: 
  [['5 PM', 'Wednesday', '4e055891fa767637fd1db8e7'], ['1 AM', 'Friday', '43a26468f964a520322c1fe3'], ['5 AM', 'Saturday', '46c77917f964a5201a4a1fe3']]
  <target_stay>: '<next_place_id>'                 ",
"output": {
    "prediction": [
        "4fd93beeb634312a5bc2ca50",
        "4f451eff19836ed001be34fe",
        "4a3f1f6af964a52095a11fe3",
        "4e81e4b65c5cb93d42acb10f",
        "4e81a1509adf897ae2fbed4f"
    ],
    "reason": "Based on the historical stays and context stays, we can observe that the user frequently visits certain places at specific times. Additionally, considering the temporal information provided in the target stay, it is likely that the user will visit places that align with their previous patterns. Therefore, the predicted next places are the ones that have been visited frequently during similar time periods and days of the week as indicated in the historical and context stays."
}
    \end{lstlisting} \\
    \hline
\end{longtable}

\newpage
\section{LLMs Explainantion Examples}
\label{app:Explanations}

\begin{table}[H]
\caption{Explanation provided by GPT-4o}
\centering
\begin{tabular}{@{}p{\textwidth}@{}}
\toprule
\textbf{Explanation GPT-4o:}\\
\\
The user has frequently visited the places '4d48864b551db1f722e9a330', \\ '4a5dd98ef964a520c2bd1fe3', '3fd66200f964a520def11ee3', '42ec1480f964a5209e261fe3', and \\ '4a3d735cf964a5204ba21fe3'. In addition, context locations provide recent activities at places \\ '4a5dd98ef964a520c2bd1fe3', '3fd66200f964a520def11ee3', and '4a3d735cf964a5204ba21fe3'. \\ It is likely that the user will visit one of these places again.
\\
\bottomrule
\end{tabular}
\label{tab:GPT-3.5_Explanation}
\end{table}

\begin{table}[H]
\caption{Explanation provided by GPT-4}
\centering
\begin{tabular}{@{}p{\textwidth}@{}}
\toprule
\textbf{Explanation GPT-4:}\\
\\
Based on the historical stays, the user has frequently visited the places with place IDs 4b7ecc2cf964a520420030e3, 41102700f964a520d60b1fe3, 4d531225f1b4a09039dee672, 4b705cd2f964a52052152de3, and 4c176109f693b7139f0f5f54. These places have been visited during different times of the day and on different days of the week. Considering the current context stays and the temporal information of the target stay, it is likely that the user will visit one of these places as his/her next location.
\\
\bottomrule
\end{tabular}
\label{tab:GPT-3.5_Explanation}
\end{table}

\begin{table}[H]
\caption{Explanation provided by GPT-3.5}
\centering
\begin{tabular}{@{}p{\textwidth}@{}}
\toprule
\textbf{Explanation GPT-3.5:}\\
\\
Based on the historical locations, the user has frequently visited places '4d48864b551db1f722e9a330', \\ '4a5dd98ef964a520c2bd1fe3', '3fd66200f964a520def11ee3', '42ec1480f964a5209e261fe3', and \\ '4a3d735cf964a5204ba21fe3'. Additionally, the context locations provide recent activities at places \\ '4a5dd98ef964a520c2bd1fe3', '3fd66200f964a520def11ee3', and '4a3d735cf964a5204ba21fe3'. \\ It is likely that the user will visit one of these places again.\\
\bottomrule
\end{tabular}
\label{tab:GPT-3.5_Explanation}
\end{table}

\begin{table}[H]
\caption{Explanation provided by Llama 3.1 8B}
\centering
\begin{tabular}{@{}p{\textwidth}@{}}
\toprule
\textbf{Explanation Llama 3.1 8B:}\\
\\
Considering historical visits, user will likely visit recurrent places.\\
\bottomrule
\end{tabular}
\label{tab:Llama 2_chat70b_Explanation}
\end{table}

\begin{table}[H]
\caption{Explanation provided by Llama 3 70B Instruct }
\centering
\begin{tabular}{@{}p{\textwidth}@{}}
\toprule
\textbf{Explanation Llama 3 70B Instruct :}\\
\\
Based on the user's historical stays, it is likely that they will visit a place they have been to before. The top five predicted places are all locations that frequently visited, which suggests a pattern of regular activity at these times\\
\bottomrule
\end{tabular}
\label{tab:Llama 2_chat70b_Explanation}
\end{table}

\begin{table}[H]
\caption{Explanation provided by Llama 3 8B Instruct }
\centering
\begin{tabular}{@{}p{\textwidth}@{}}
\toprule
\textbf{Explanation Llama 3 8B Instruct :}\\
\\
Based on the user's historical stays and contextual information, it is likely that the user will visit a place similar to previous destinations.\\
\bottomrule
\end{tabular}
\label{tab:Llama 2_chat70b_Explanation}
\end{table}

\begin{table}[H]
\caption{Explanation provided by Llama 3 70B}
\centering
\begin{tabular}{@{}p{\textwidth}@{}}
\toprule
\textbf{Explanation Llama 3 70B:}\\
\\
Contextural visits are recurrent and historical visits indicate the user will likely visits similar locations.\\
\bottomrule
\end{tabular}
\label{tab:Llama 2_chat70b_Explanation}
\end{table}

\begin{table}[H]
\caption{Explanation provided by Llama 3 8B}
\centering
\begin{tabular}{@{}p{\textwidth}@{}}
\toprule
\textbf{Explanation Llama 3 8B:}\\
\\
Based on the user's historical stays and contextual information, it is likely that the user will visit a place similar to previous destinations\\
\bottomrule
\end{tabular}
\label{tab:Llama 2_chat70b_Explanation}
\end{table}

\begin{table}[H]
\caption{Explanation provided by Llama 2 Chat 70B}
\centering
\begin{tabular}{@{}p{\textwidth}@{}}
\toprule
\textbf{Explanation Llama 2 Chat 70B:}\\
\\
Based on the user's historical stays, it is likely that they will visit one of the places they have previously been. The top five predicted places are all locations that the user has visited before at around the same time.\\
\bottomrule
\end{tabular}
\label{tab:Llama 2_chat70b_Explanation}
\end{table}

\begin{table}[H]
\caption{Explanation provided by Llama 2 Chat 13B}
\centering
\begin{tabular}{@{}p{\textwidth}@{}}
\toprule
\textbf{Explanation Llama 2 Chat 13B:}\\
\\
The predicted places are based on the user's historical stays, contextual information from recent stays, and temporal information. The top three predictions are all locations that the user has visited before\\
\bottomrule
\end{tabular}
\label{tab:Llama 2_chat13b_Explanation}
\end{table}

\begin{table}[H]
\caption{Explanation provided by Llama 2 Chat 7B}
\centering
\begin{tabular}{@{}p{\textwidth}@{}}
\toprule
\textbf{Explanation Llama 2 Chat 7B:}\\
\\
The user's activity pattern and recent contextual information suggest that they are likely to visit places that are similar to their previous locations. For example, they have repeatedly visited a coffee \\ shop near their workplace during lunchtime.\\
\bottomrule
\end{tabular}
\label{tab:Llama 2_chat7b_Explanation}
\end{table}

\begin{table}[H]
\caption{Explanation provided by Llama 2 70B}
\centering
\begin{tabular}{@{}p{\textwidth}@{}}
\toprule
\textbf{Explanation Llama 2 70B:}\\
\\
Based on the user ’ s historical stays ,
it is likely that they will visit places with high repeat
visits \\
\bottomrule
\end{tabular}
\label{tab:Llama 2_70b_Explanation}
\end{table}

\begin{table}[H]
\caption{Explanation provided by Llama 2 13B}
\centering
\begin{tabular}{@{}p{\textwidth}@{}}
\toprule
\textbf{Explanation Llama 2 13B:}\\
\\
User will likely visit places with
high repeat visits \\
\bottomrule
\end{tabular}
\label{tab:Llama 2_13b_Explanation}
\end{table}

\begin{table}[H]
\caption{Explanation provided by Llama 2 7B}
\centering
\begin{tabular}{@{}p{\textwidth}@{}}
\toprule
\textbf{Explanation Llama 2 7B:}\\
\\
User will likely visit places with
high repeat visits \\
\bottomrule
\end{tabular}
\label{tab:Llama 2_7b_Explanation}
\end{table}

\begin{table}[H]
\caption{Explanation provided by Mistral 7B}
\centering
\begin{tabular}{@{}p{\textwidth}@{}}
\toprule
\textbf{Explanation Mistral 7B:}\\
\\
The user will likely visit one of these places
again  \\
\bottomrule
\end{tabular}
\label{tab:Mistral 7B_Explanation}
\end{table}

\newpage
\section{ACC@1 of LLMs}
\label{app:acc_1}
\begin{table}[H]
\centering
\caption{ACC@1 of selected LLMs prompted to perform zero-shot, one-shot and 3-shot next location prediction on all the datasets.}
\resizebox{0.8\textwidth}{!}{%
\begin{tabular}{l|lll|lll|lll|}
          & \multicolumn{3}{c|}{F. NYC}                   & \multicolumn{3}{c|}{F. TKY}                   & \multicolumn{3}{c|}{Our}                      \\ 
          & zero-shot             & one-shot             & few-shot             & zero-shot             & one-shot             & few-shot             & zero-shot             & one-shot             & few-shot             \\ \hline
Llama 2 70B chat        & 0.179 & 0.096 & 0.146 & 0.199 & 0.127 & 0.158 & 0.122 & 0.116 & 0.108 \\
Llama 2 13B chat        & 0.099 & 0.083 & 0.029 & 0.078 & 0.061 & 0.007 & 0.103 & 0.095 & 0.029 \\
Llama 2 7B chat         & 0.086 & 0.057 & 0.021 & 0.047 & 0.051 & 0.006 & 0.088 & 0.082 & 0.025 \\
Llama 2 70B             & 0.062 & 0.078 & 0.091 & 0.042 & 0.038 & 0.041 & 0.056 & 0.048 & 0.061 \\
Llama 2 13B             & 0.011 & 0.010 & 0.006 & 0.009 & 0.011 & 0.004 & 0.033 & 0.035 & 0.028 \\
Llama 2 7B              & 0.009 & 0.005 & 0.003 & 0.005 & 0.006 & 0.001 & 0.010 & 0.006 & 0.001 \\
GPT-3.5                 & 0.192 & 0.189 & 0.196 & 0.199 & 0.213 & 0.207 & 0.184 & 0.195 & 0.189 \\
GPT-4                   & 0.195 & 0.190 & 0.211 & 0.205 & 0.226 & 0.208 & 0.188 & 0.193 & 0.189 \\
GPT-4o                  & 0.199 & 0.198 & 0.220 & 0.209 & 0.237 & 0.211 & 0.202 & 0.206 & 0.200 \\
Llama 3 8B              & 0.011 & 0.012 & 0.011 & 0.009 & 0.009 & 0.008 & 0.009 & 0.012 & 0.012 \\
Llama 3 70B             & 0.084 & 0.091 & 0.106 & 0.077 & 0.081 & 0.055 & 0.063 & 0.071 & 0.070 \\
Llama 3 8B Instruct     & 0.101 & 0.085 & 0.027 & 0.081 & 0.078 & 0.011 & 0.107 & 0.107 & 0.049 \\
Llama 3 70B Instruct    & 0.196 & 0.201 & 0.225 & 0.199 & 0.228 & 0.218 & 0.205 & 0.212 & 0.208 \\
Llama 3.1 8B            & 0.019 & 0.019 & 0.015 & 0.010 & 0.008 & 0.010 & 0.007 & 0.010 & 0.009 \\
Mistral 7B              & 0.087 & 0.079 & 0.030 & 0.069 & 0.058 & 0.009 & 0.088 & 0.054 & 0.033 \\
\hline
\end{tabular}%
}
\label{tab:acc_1}
\end{table}

\section{ACC@3 of LLMs}
\label{app:acc_3}
\begin{table}[H]
\centering
\caption{ACC@3 of selected LLMs prompted to perform zero-shot, one-shot and 3-shot next location prediction on all the datasets.}
\resizebox{0.8\textwidth}{!}{%
\begin{tabular}{l|lll|lll|lll|}
          & \multicolumn{3}{c|}{F. NYC}                   & \multicolumn{3}{c|}{F. TKY}                   & \multicolumn{3}{c|}{Our}                      \\ \hline
          & zero-shot             & one-shot             & few-shot             & zero-shot             & one-shot             & few-shot             & zero-shot             & one-shot             & few-shot             \\ 
Llama 2 70B chat        & 0.258 & 0.188 & 0.224 & 0.278 & 0.200 & 0.236 & 0.217 & 0.192 & 0.199 \\
Llama 2 13B chat        & 0.144 & 0.135 & 0.026 & 0.162 & 0.155 & 0.034 & 0.188 & 0.176 & 0.055 \\
Llama 2 7B chat         & 0.115 & 0.173 & 0.061 & 0.122 & 0.152 & 0.061 & 0.119 & 0.168 & 0.063 \\
Llama 2 70B             & 0.158 & 0.166 & 0.177 & 0.107 & 0.120 & 0.131 & 0.142 & 0.138 & 0.145 \\
Llama 2 13B             & 0.099 & 0.116 & 0.088 & 0.099 & 0.103 & 0.077 & 0.072 & 0.099 & 0.077 \\
Llama 2 7B              & 0.068 & 0.066 & 0.006 & 0.072 & 0.071 & 0.008 & 0.056 & 0.044 & 0.005 \\
GPT-3.5                 & 0.282 & 0.293 & 0.289 & 0.289 & 0.311 & 0.298 & 0.266 & 0.282 & 0.265 \\
GPT-4                   & 0.286 & 0.298 & 0.292 & 0.298 & 0.292 & 0.312 & 0.285 & 0.299 & 0.298 \\
GPT-4o                  & 0.287 & 0.298 & 0.299 & 0.302 & 0.304 & 0.295 & 0.280 & 0.292 & 0.293 \\
Llama 3 8B              & 0.082 & 0.081 & 0.011 & 0.091 & 0.077 & 0.010 & 0.082 & 0.066 & 0.009 \\
Llama 3 70B             & 0.117 & 0.112 & 0.119 & 0.122 & 0.118 & 0.119 & 0.138 & 0.119 & 0.125 \\
Llama 3 8B Instruct     & 0.148 & 0.159 & 0.096 & 0.142 & 0.166 & 0.075 & 0.172 & 0.188 & 0.099 \\
Llama 3 70B Instruct    & 0.309 & 0.276 & 0.311 & 0.322 & 0.249 & 0.318 & 0.263 & 0.233 & 0.267 \\
Llama 3.1 8B            & 0.096 & 0.098 & 0.059 & 0.098 & 0.095 & 0.081 & 0.089 & 0.092 & 0.074 \\
Mistral 7B              & 0.136 & 0.149 & 0.162 & 0.136 & 0.155 & 0.158 & 0.161 & 0.165 & 0.164 \\
\hline
\end{tabular}%
}
\label{tab:acc_3}
\end{table}

\newpage
\section{ACC@5 of LLMs}
\label{app:acc_5}
\begin{table}[h!]
\centering
\caption{ACC@5 of selected LLMs prompted to perform zero-shot, one-shot and 3-shot next location prediction on all the datasets.}
\resizebox{0.8\textwidth}{!}{%
\begin{tabular}{l|lll|lll|lll|}
              & \multicolumn{3}{c|}{F. NYC}                   & \multicolumn{3}{c|}{F. TKY}                   & \multicolumn{3}{c|}{Our}                      \\
         & zero-shot             & one-shot             & few-shot             & zero-shot             & one-shot             & few-shot             & zero-shot             & one-shot             & few-shot             \\ \hline
Llama 2 70B chat & 0.305 & 0.228 & 0.281 & 0.291 & 0.214 & 0.250 & 0.231 & 0.205 & 0.219 \\
Llama 2 13B chat & 0.172 & 0.168 & 0.047 & 0.169 & 0.162 & 0.039 & 0.190 & 0.181 & 0.060 \\
Llama 2 7B chat & 0.137 & 0.185 & 0.066 & 0.136 & 0.178 & 0.069 & 0.135 & 0.182 & 0.064 \\
Llama 2 70B & 0.163 & 0.172 & 0.183 & 0.126 & 0.139 & 0.144 & 0.148 & 0.149 & 0.156 \\
Llama 2 13B & 0.105 & 0.126 & 0.092 & 0.116 & 0.122 & 0.089 & 0.089 & 0.105 & 0.082 \\
Llama 2 7B & 0.077 & 0.071 & 0.006 & 0.084 & 0.078 & 0.012 & 0.065 & 0.059 & 0.007 \\
GPT-3.5 & 0.298 & 0.310 & 0.302 & 0.316 & 0.347 & 0.323 & 0.280 & 0.291 & 0.279 \\
GPT-4 & 0.307 & 0.312 & 0.306 & 0.311 & 0.316 & 0.312 & 0.285 & 0.299 & 0.298 \\
GPT-4o & 0.309 & 0.317 & 0.311 & 0.316 & 0.329 & 0.315 & 0.288 & 0.304 & 0.302 \\
Llama 3 8B & 0.100 & 0.094 & 0.014 & 0.108 & 0.095 & 0.014 & 0.091 & 0.088 & 0.011 \\
Llama 3 70B & 0.148 & 0.133 & 0.139 & 0.146 & 0.135 & 0.138 & 0.156 & 0.139 & 0.143 \\
Llama 3 8B Instruct & 0.165 & 0.196 & 0.128 & 0.154 & 0.188 & 0.099 & 0.189 & 0.209 & 0.112 \\
Llama 3 70B Instruct & 0.338 & 0.287 & 0.322 & 0.341 & 0.266 & 0.329 & 0.287 & 0.257 & 0.281 \\
Llama 3.1 8B & 0.110 & 0.112 & 0.087 & 0.113 & 0.112 & 0.094 & 0.107 & 0.109 & 0.099 \\
Mistral 7B & 0.156 & 0.165 & 0.171 & 0.159 & 0.168 & 0.173 & 0.178 & 0.182 & 0.188 \\

\hline
\end{tabular}%
}
\label{tab:acc_zero_shot}
\end{table}

\newpage
\section{Context and Historical Role}
\label{app:ayblation}

\begin{table}[h]
\centering
\caption{Relative improvements of selected LLMs evaluated on different datasets (Ferrara, New York, Tokyo) under varying configurations of context (C) and historical stays (H). The standard configuration (C=6, H=15) is shown in the first column for reference.}
\resizebox{0.6\textwidth}{!}{%
\begin{tabular}{llrrrrrrr}
\toprule
Model Name & Dataset & C=6; H=15 & C=0 & C=3 & C=12 & H=0 & H=7 & H=30 \\
\midrule
GPT-3.5 & Ferrara & 0.280 & -0.261 & -0.121 & -0.025 & -0.389 & -0.100 & 0.032 \\
GPT-3.5 & New York & 0.298 & -0.144 & -0.070 & 0.050 & -0.695 & -0.060 & 0.168 \\
GPT-3.5 & Tokyo & 0.316 & -0.101 & -0.092 & 0.051 & -0.699 & -0.063 & 0.114 \\
\midrule
GPT-4 & Ferrara & 0.285 & -0.274 & -0.154 & -0.028 & -0.368 & -0.070 & 0.042 \\
GPT-4 & New York & 0.307 & -0.029 & -0.016 & 0.029 & -0.661 & -0.072 & 0.091 \\
GPT-4 & Tokyo & 0.311 & -0.023 & -0.006 & 0.035 & -0.666 & -0.061 & 0.087 \\
\midrule
GPT-4o & Ferrara & 0.288 & -0.271 & -0.153 & -0.024 & -0.351 & -0.076 & 0.045 \\
GPT-4o & New York & 0.309 & -0.026 & -0.016 & 0.042 & -0.641 & -0.006 & 0.126 \\
GPT-4o & Tokyo & 0.316 & -0.009 & 0.000 & 0.051 & -0.623 & -0.016 & 0.146 \\
\midrule
Llama 3 70B  & Ferrara & 0.156 & -0.218 & -0.058 & 0.019 & -0.237 & -0.071 & 0.000 \\
Llama 3 70B  & New York & 0.148 & -0.230 & -0.081 & -0.061 & -0.209 & -0.095 & 0.007 \\
Llama 3 70B  & Tokyo & 0.146 & -0.205 & -0.096 & -0.103 & -0.260 & -0.110 & 0.014 \\
\midrule
Llama 3 70B Instruct  & Ferrara & 0.287 & -0.265 & -0.111 & -0.021 & -0.341 & -0.070 & 0.049 \\
Llama 3 70B Instruct  & New York & 0.338 & -0.107 & -0.083 & 0.015 & -0.689 & -0.115 & 0.041 \\
Llama 3 70B Instruct  & Tokyo & 0.341 & -0.097 & -0.088 & 0.021 & -0.680 & -0.126 & 0.059 \\
\midrule
Llama 3 8B  & Ferrara & 0.091 & -0.396 & -0.242 & 0.176 & -0.582 & 0.121 & 0.418 \\
Llama 3 8B  & New York & 0.100 & -0.410 & -0.260 & 0.090 & -0.490 & 0.010 & 0.260 \\
Llama 3 8B  & Tokyo & 0.108 & -0.435 & -0.269 & 0.046 & -0.500 & 0.009 & 0.130 \\
\midrule
Llama 3 8B  Instruct & Ferrara & 0.189 & -0.228 & -0.042 & 0.053 & -0.429 & -0.069 & 0.005 \\
Llama 3 8B  Instruct & New York & 0.165 & -0.339 & -0.327 & 0.127 & -0.461 & -0.206 & 0.073 \\
Llama 3 8B  Instruct & Tokyo & 0.154 & -0.208 & -0.052 & 0.162 & -0.357 & -0.045 & 0.351 \\
\midrule
Llama 3,1 8B & Ferrara & 0.107 & -0.093 & -0.056 & 0.103 & -0.542 & 0.000 & 0.243 \\
Llama 3,1 8B & New York & 0.110 & -0.055 & -0.018 & 0.100 & -0.555 & -0.018 & 0.236 \\
Llama 3,1 8B & Tokyo & 0.113 & -0.062 & -0.035 & 0.097 & -0.540 & -0.035 & 0.248 \\
\midrule
Llama 2 13B & Ferrara & 0.089 & -0.348 & -0.191 & 0.213 & -0.539 & 0.169 & 0.517 \\
Llama 2 13B & New York & 0.105 & -0.352 & -0.171 & 0.067 & -0.457 & -0.038 & 0.200 \\
Llama 2 13B & Tokyo & 0.116 & -0.405 & -0.147 & 0.017 & -0.517 & -0.034 & 0.069 \\
\midrule
Llama 2 Chat 13B & Ferrara & 0.190 & -0.174 & -0.032 & 0.068 & -0.305 & -0.021 & 0.021 \\
Llama 2 Chat 13B & New York & 0.172 & -0.267 & -0.192 & 0.128 & -0.424 & -0.099 & 0.151 \\
Llama 2 Chat 13B & Tokyo & 0.169 & -0.249 & -0.065 & 0.071 & -0.379 & -0.077 & 0.325 \\
\midrule
Llama 2 70B & Ferrara & 0.148 & -0.318 & -0.270 & 0.061 & -0.446 & -0.041 & 0.122 \\
Llama 2 70B & New York & 0.163 & -0.252 & -0.123 & 0.049 & -0.411 & 0.000 & 0.153 \\
Llama 2 70B & Tokyo & 0.126 & -0.302 & -0.262 & 0.079 & -0.548 & -0.024 & 0.016 \\
\midrule
Llama 2 Chat 70B & Ferrara & 0.231 & -0.203 & -0.052 & 0.052 & -0.320 & -0.013 & 0.026 \\
Llama 2 Chat 70B & New York & 0.305 & -0.154 & -0.138 & 0.049 & -0.718 & -0.066 & 0.062 \\
Llama 2 Chat 70B & Tokyo & 0.291 & -0.144 & -0.107 & 0.027 & -0.694 & -0.131 & 0.069 \\
\midrule
Llama 2 7B & Ferrara & 0.065 & -0.585 & -0.215 & 0.046 & -0.877 & -0.154 & 0.277 \\
Llama 2 7B & New York & 0.077 & -0.701 & -0.247 & 0.078 & -0.792 & -0.195 & 0.143 \\
Llama 2 7B & Tokyo & 0.084 & -0.714 & -0.274 & 0.024 & -0.821 & -0.036 & 0.167 \\
\midrule
Llama 2 Chat 7B & Ferrara & 0.135 & -0.385 & -0.170 & 0.037 & -0.585 & -0.104 & 0.156 \\
Llama 2 Chat 7B & New York & 0.137 & -0.292 & -0.117 & 0.182 & -0.606 & -0.022 & 0.569 \\
Llama 2 Chat 7B & Tokyo & 0.136 & -0.353 & -0.096 & 0.162 & -0.654 & -0.074 & 0.603 \\
\midrule
Mistral 7B & Ferrara & 0.178 & -0.112 & -0.017 & 0.163 & -0.337 & -0.022 & 0.129 \\
Mistral 7B & New York & 0.156 & -0.160 & -0.122 & 0.276 & -0.468 & 0.038 & 0.340 \\
Mistral 7B & Tokyo & 0.159 & -0.182 & -0.145 & 0.358 & -0.384 & 0.063 & 0.371 \\
\bottomrule
\label{tab:model_data_variation_ordered}
\end{tabular}
}
\end{table}

\begin{table}[h]
\centering
\caption{ACC@5 of selected LLMs followed by their respective Standard Deviations (SD), both evaluated on different datasets under varying historical (H) and contextual (C) configurations.}
\resizebox{\textwidth}{!}{%

\begin{tabular}{p{3.8 cm} p{1.8 cm} p{2.5 cm} p{1.1cm} p{1.2cm} p{1.1cm} p{1.2cm} p{1.1cm} p{1.2cm} p{1.1cm} p{1.2cm} p{1.1cm} p{1.2cm} p{1.1cm} p{1.2cm}}
\toprule
Model Name & Dataset & C=6; H=15 & C = 0 & SD & C = 3 & SD & C = 12 & SD & H = 0 & SD & H = 7 & SD & H = 30 & SD \\
\midrule
GPT-3.5 & Ferrara & 0.280 & 0.207 & 0.008 & 0.246 & 0.011 & 0.273 & 0.008 & 0.171 & 0.011 & 0.252 & 0.007 & 0.289 & 0.014 \\
GPT-3.5 & New York & 0.298 & 0.255 & 0.007 & 0.277 & 0.013 & 0.313 & 0.013 & 0.091 & 0.007 & 0.280 & 0.011 & 0.348 & 0.008 \\
GPT-3.5 & Tokyo & 0.316 & 0.284 & 0.012 & 0.287 & 0.001 & 0.332 & 0.014 & 0.095 & 0.007 & 0.296 & 0.011 & 0.352 & 0.012 \\
\midrule
GPT-4 & Ferrara & 0.285 & 0.207 & 0.011 & 0.241 & 0.007 & 0.277 & 0.009 & 0.180 & 0.012 & 0.265 & 0.015 & 0.297 & 0.016 \\
GPT-4 & New York & 0.307 & 0.298 & 0.016 & 0.302 & 0.012 & 0.316 & 0.013 & 0.104 & 0.011 & 0.285 & 0.008 & 0.335 & 0.001 \\
GPT-4 & Tokyo & 0.311 & 0.304 & 0.014 & 0.309 & 0.012 & 0.322 & 0.013 & 0.104 & 0.009 & 0.292 & 0.009 & 0.338 & 0.011 \\
\midrule
GPT-4o & Ferrara & 0.288 & 0.210 & 0.013 & 0.244 & 0.001 & 0.281 & 0.015 & 0.187 & 0.008 & 0.266 & 0.008 & 0.301 & 0.013 \\
GPT-4o & New York & 0.309 & 0.301 & 0.011 & 0.304 & 0.008 & 0.322 & 0.015 & 0.111 & 0.008 & 0.307 & 0.008 & 0.348 & 0.009 \\
GPT-4o & Tokyo & 0.316 & 0.313 & 0.007 & 0.316 & 0.014 & 0.332 & 0.008 & 0.119 & 0.012 & 0.311 & 0.001 & 0.362 & 0.015 \\
\midrule
Llama 3 70B  & Ferrara & 0.156 & 0.122 & 0.009 & 0.147 & 0.014 & 0.159 & 0.012 & 0.119 & 0.011 & 0.145 & 0.016 & 0.156 & 0.009 \\
Llama 3 70B  & New York & 0.148 & 0.114 & 0.015 & 0.136 & 0.016 & 0.139 & 0.016 & 0.117 & 0.014 & 0.134 & 0.014 & 0.149 & 0.013 \\
Llama 3 70B  & Tokyo & 0.146 & 0.116 & 0.012 & 0.132 & 0.007 & 0.131 & 0.006 & 0.108 & 0.011 & 0.130 & 0.015 & 0.148 & 0.001 \\
\midrule
Llama 3 70B Instruct  & Ferrara & 0.287 & 0.211 & 0.011 & 0.255 & 0.015 & 0.281 & 0.008 & 0.189 & 0.009 & 0.267 & 0.008 & 0.301 & 0.012 \\
Llama 3 70B Instruct  & New York & 0.338 & 0.302 & 0.015 & 0.31 & 0.015 & 0.343 & 0.012 & 0.105 & 0.016 & 0.299 & 0.008 & 0.352 & 0.008 \\
Llama 3 70B Instruct  & Tokyo & 0.341 & 0.308 & 0.013 & 0.311 & 0.011 & 0.348 & 0.008 & 0.109 & 0.013 & 0.298 & 0.009 & 0.361 & 0.012 \\
\midrule
Llama 3 8B  & Ferrara & 0.091 & 0.055 & 0.012 & 0.069 & 0.001 & 0.107 & 0.015 & 0.038 & 0.002 & 0.102 & 0.021 & 0.129 & 0.012 \\
Llama 3 8B  & New York & 0.1 & 0.059 & 0.011 & 0.074 & 0.002 & 0.109 & 0.001 & 0.051 & 0.006 & 0.101 & 0.013 & 0.126 & 0.007 \\
Llama 3 8B  & Tokyo & 0.108 & 0.061 & 0.007 & 0.079 & 0.009 & 0.113 & 0.009 & 0.054 & 0.001 & 0.109 & 0.012 & 0.122 & 0.011 \\
\midrule
Llama 3 8B  Instruct & Ferrara & 0.189 & 0.146 & 0.015 & 0.181 & 0.007 & 0.199 & 0.007 & 0.108 & 0.008 & 0.176 & 0.012 & 0.19 & 0.014 \\
Llama 3 8B  Instruct & New York & 0.165 & 0.109 & 0.007 & 0.111 & 0.011 & 0.186 & 0.011 & 0.089 & 0.012 & 0.131 & 0.014 & 0.177 & 0.001 \\
Llama 3 8B  Instruct & Tokyo & 0.154 & 0.122 & 0.008 & 0.146 & 0.008 & 0.179 & 0.007 & 0.099 & 0.016 & 0.147 & 0.013 & 0.208 & 0.001 \\
\midrule
Llama 3.1 8B & Ferrara & 0.107 & 0.097 & 0.009 & 0.101 & 0.011 & 0.118 & 0.012 & 0.049 & 0.007 & 0.107 & 0.013 & 0.133 & 0.013 \\
Llama 3.1 8B & New York & 0.11 & 0.104 & 0.011 & 0.108 & 0.008 & 0.121 & 0.007 & 0.049 & 0.007 & 0.108 & 0.009 & 0.136 & 0.008 \\
Llama 3.1 8B & Tokyo & 0.113 & 0.106 & 0.011 & 0.109 & 0.001 & 0.124 & 0.012 & 0.052 & 0.011 & 0.109 & 0.008 & 0.141 & 0.013 \\
\midrule
Llama 2 13B & Ferrara & 0.089 & 0.058 & 0.001 & 0.072 & 0.009 & 0.108 & 0.014 & 0.041 & 0.016 & 0.104 & 0.015 & 0.135 & 0.011 \\
Llama 2 13B & New York & 0.105 & 0.068 & 0.013 & 0.087 & 0.001 & 0.112 & 0.012 & 0.057 & 0.013 & 0.101 & 0.015 & 0.126 & 0.001 \\
Llama 2 13B & Tokyo & 0.116 & 0.069 & 0.007 & 0.099 & 0.014 & 0.118 & 0.001 & 0.056 & 0.016 & 0.112 & 0.007 & 0.124 & 0.012 \\
\midrule
Llama 2 Chat 13B & Ferrara & 0.19 & 0.157 & 0.014 & 0.184 & 0.016 & 0.203 & 0.008 & 0.132 & 0.014 & 0.186 & 0.014 & 0.194 & 0.016 \\
Llama 2 Chat 13B & New York & 0.172 & 0.126 & 0.007 & 0.139 & 0.015 & 0.194 & 0.001 & 0.099 & 0.012 & 0.155 & 0.007 & 0.198 & 0.016 \\
Llama 2 Chat 13B & Tokyo & 0.169 & 0.127 & 0.015 & 0.158 & 0.009 & 0.181 & 0.013 & 0.105 & 0.016 & 0.156 & 0.007 & 0.224 & 0.001 \\
\midrule
Llama 2 70B & Ferrara & 0.148 & 0.101 & 0.013 & 0.108 & 0.011 & 0.157 & 0.009 & 0.082 & 0.009 & 0.142 & 0.015 & 0.166 & 0.012 \\
Llama 2 70B & New York & 0.163 & 0.122 & 0.015 & 0.143 & 0.014 & 0.171 & 0.016 & 0.096 & 0.001 & 0.163 & 0.001 & 0.188 & 0.011 \\
Llama 2 70B & Tokyo & 0.126 & 0.088 & 0.007 & 0.093 & 0.015 & 0.136 & 0.012 & 0.057 & 0.014 & 0.123 & 0.012 & 0.128 & 0.008 \\
\midrule
Llama 2 Chat 70B & Ferrara & 0.231 & 0.184 & 0.005 & 0.219 & 0.001 & 0.243 & 0.001 & 0.157 & 0.013 & 0.228 & 0.015 & 0.237 & 0.011 \\
Llama 2 Chat 70B & New York & 0.305 & 0.258 & 0.007 & 0.263 & 0.015 & 0.32 & 0.013 & 0.086 & 0.012 & 0.285 & 0.013 & 0.324 & 0.015 \\
Llama 2 Chat 70B & Tokyo & 0.291 & 0.249 & 0.012 & 0.26 & 0.001 & 0.299 & 0.008 & 0.089 & 0.014 & 0.253 & 0.011 & 0.311 & 0.007 \\
\midrule
Llama 2 7B & Ferrara & 0.065 & 0.027 & 0.008 & 0.051 & 0.001 & 0.068 & 0.001 & 0.008 & 0.001 & 0.055 & 0.012 & 0.083 & 0.015 \\
Llama 2 7B & New York & 0.077 & 0.023 & 0.001 & 0.058 & 0.001 & 0.083 & 0.001 & 0.016 & 0.001 & 0.062 & 0.011 & 0.088 & 0.013 \\
Llama 2 7B & Tokyo & 0.084 & 0.024 & 0.001 & 0.061 & 0.002 & 0.086 & 0.012 & 0.015 & 0.001 & 0.081 & 0.011 & 0.098 & 0.015 \\
\midrule
Llama 2 Chat 7B & Ferrara & 0.135 & 0.083 & 0.012 & 0.112 & 0.016 & 0.14 & 0.007 & 0.056 & 0.011 & 0.121 & 0.001 & 0.156 & 0.015 \\
Llama 2 Chat 7B & New York & 0.137 & 0.097 & 0.001 & 0.121 & 0.009 & 0.162 & 0.013 & 0.054 & 0.001 & 0.134 & 0.009 & 0.215 & 0.008 \\
Llama 2 Chat 7B & Tokyo & 0.136 & 0.088 & 0.013 & 0.123 & 0.008 & 0.158 & 0.014 & 0.047 & 0.016 & 0.126 & 0.015 & 0.218 & 0.007 \\
\midrule
Mistral 7B & Ferrara & 0.178 & 0.158 & 0.001 & 0.175 & 0.009 & 0.207 & 0.011 & 0.118 & 0.009 & 0.174 & 0.012 & 0.201 & 0.009 \\
Mistral 7B & New York & 0.156 & 0.131 & 0.008 & 0.137 & 0.009 & 0.199 & 0.012 & 0.083 & 0.001 & 0.162 & 0.015 & 0.209 & 0.008 \\
Mistral 7B & Tokyo & 0.159 & 0.130 & 0.012 & 0.136 & 0.007 & 0.216 & 0.011 & 0.098 & 0.011 & 0.169 & 0.009 & 0.218 & 0.014 \\
\bottomrule
\label{tab:model_data_with_errors}
\end{tabular}

}
\end{table}

\newpage
\section{ACC@5 Performances of Tokyo and Ferrara}

\begin{figure}[H]
    \centering
    \includegraphics[width=1\linewidth]{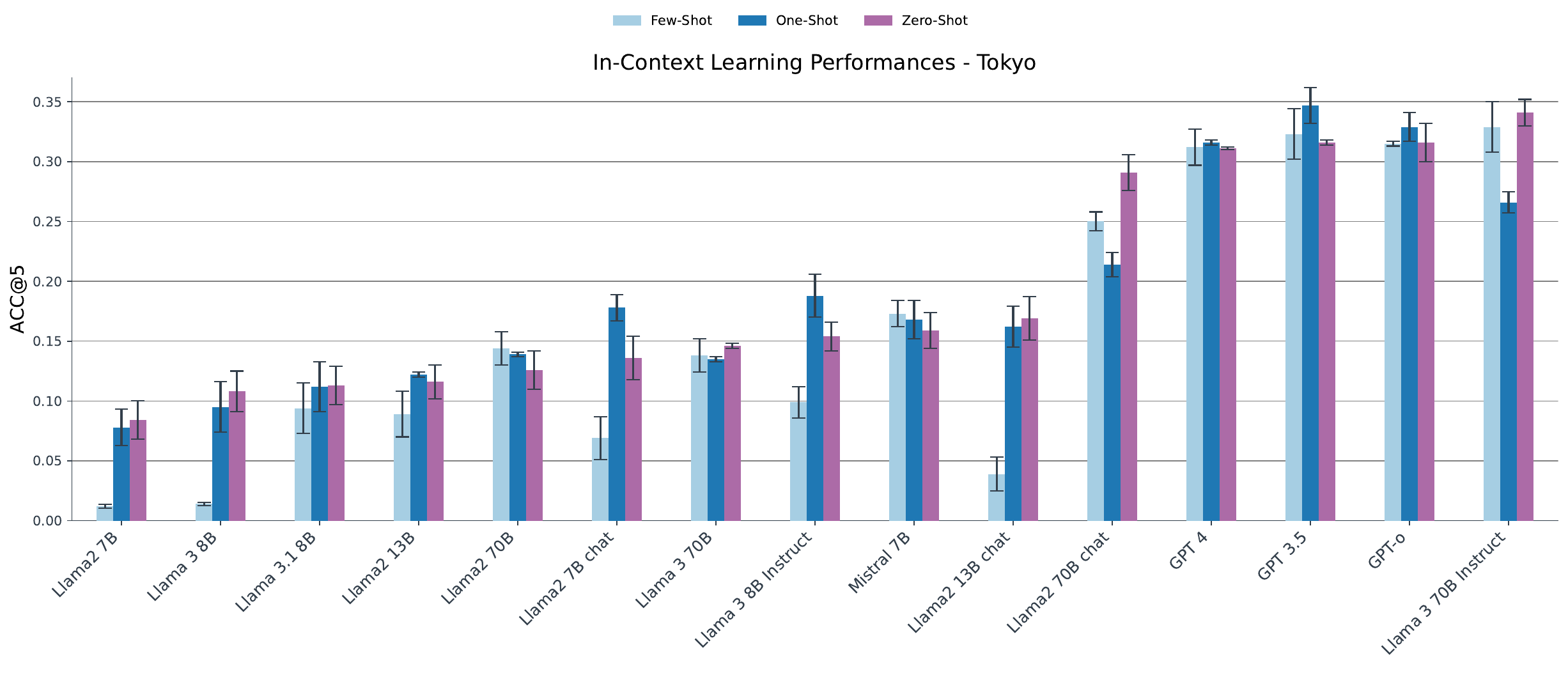}
    \caption{ACC@5 of LLMs with zero shot (purple), one shot (dark blue) and few shot (light blue) prompts in Tokyo. 
    }
    \label{fig:acc_zero_shot}
\end{figure}

\begin{figure}[H]
    \centering
    \includegraphics[width=1\linewidth]{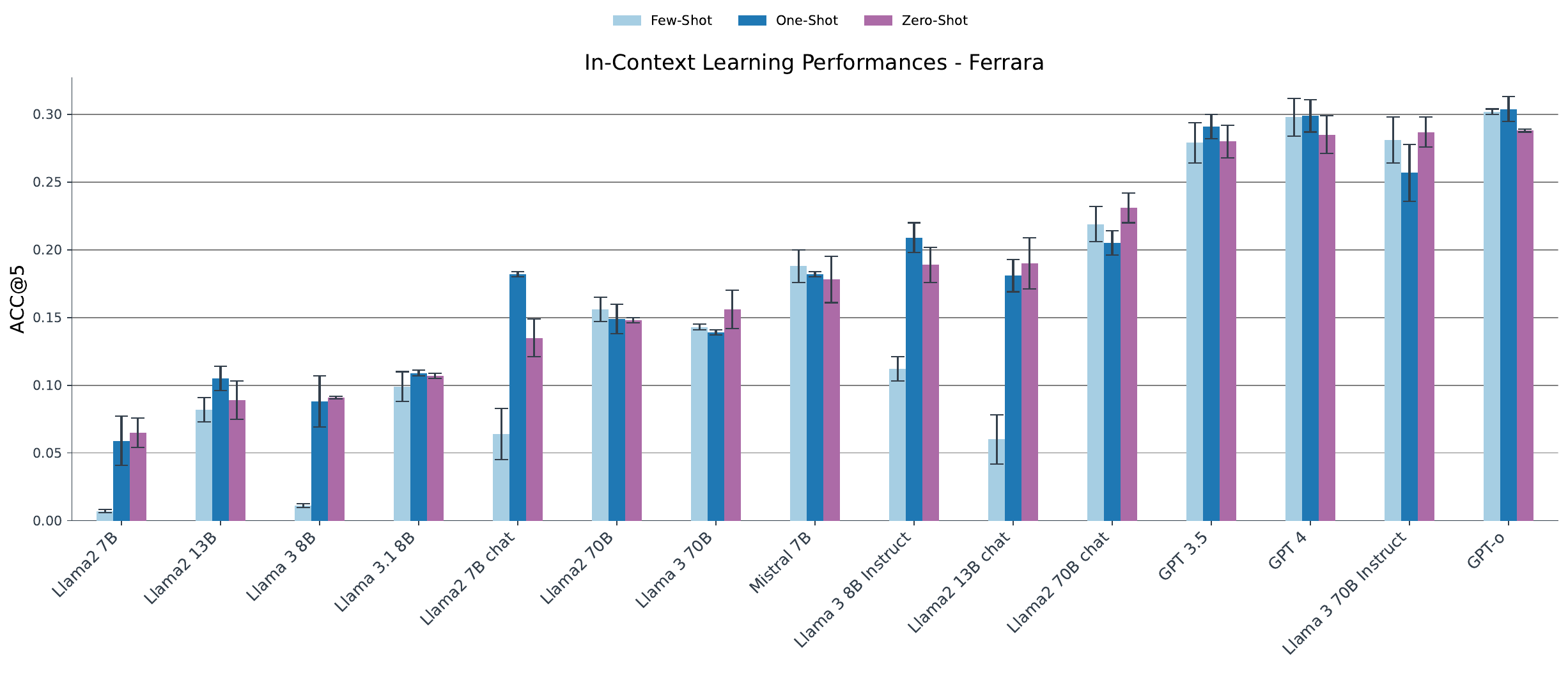}
    \caption{ACC@5 of LLMs with zero shot (purple), one shot (dark blue) and few shot (light blue) prompts in Ferrara. 
    }
    \label{fig:acc_zero_shot}
\end{figure}

\newpage
\section{Relative Improvements of Tokyo and Ferrara}

\begin{figure}[H]
    \centering
    \includegraphics[width=1\linewidth]{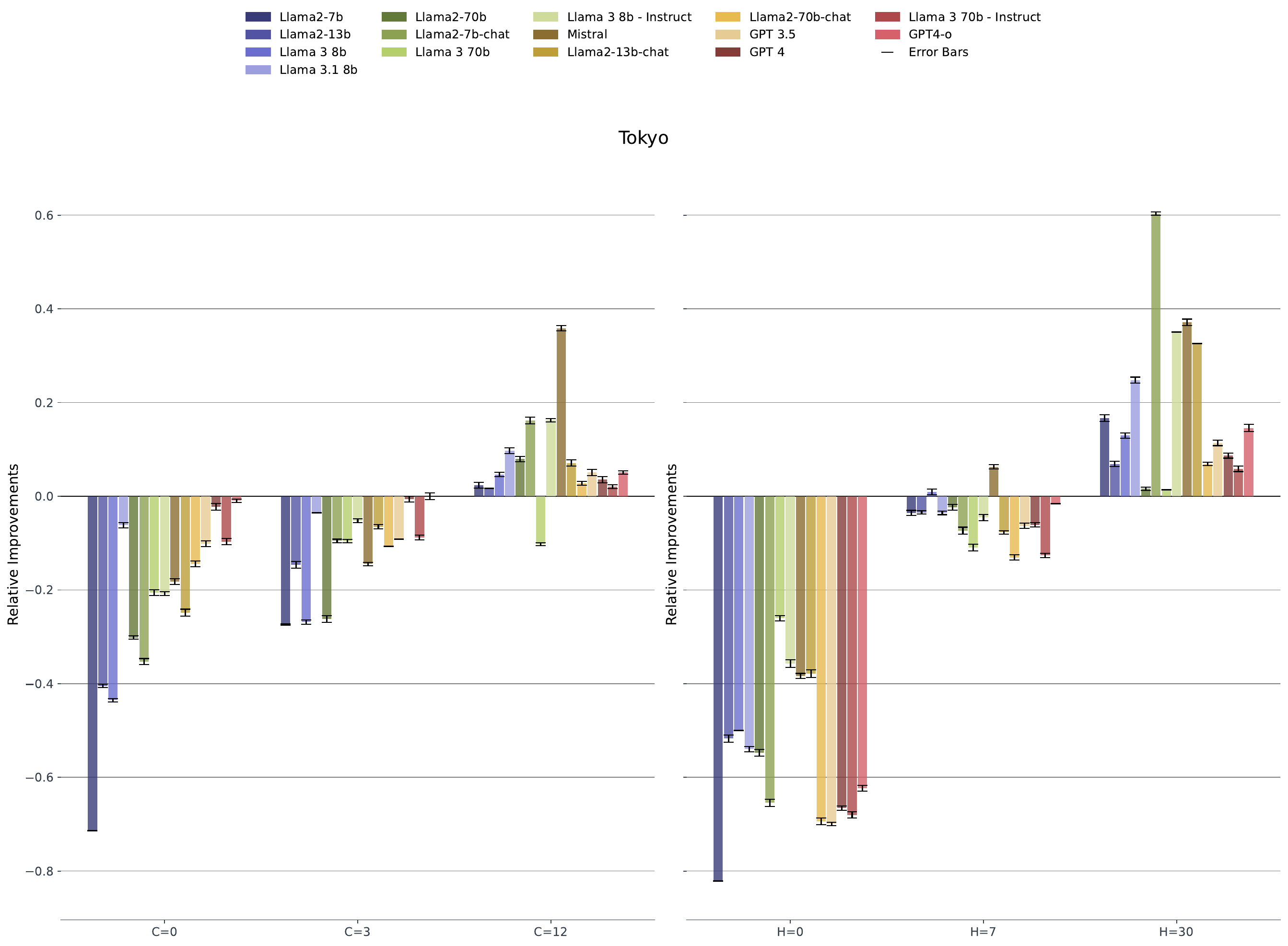}
    \caption{ Results in terms of relative improvements and drops with respect to the scenario with $C=6, H=15$ for the city of Tokyo when we modify the availability of contextual $C$ (left) and historical $H$ (right) information. }
    \label{fig:conf_c_h}
\end{figure}

\begin{figure}[H]
    \centering
    \includegraphics[width=1\linewidth]{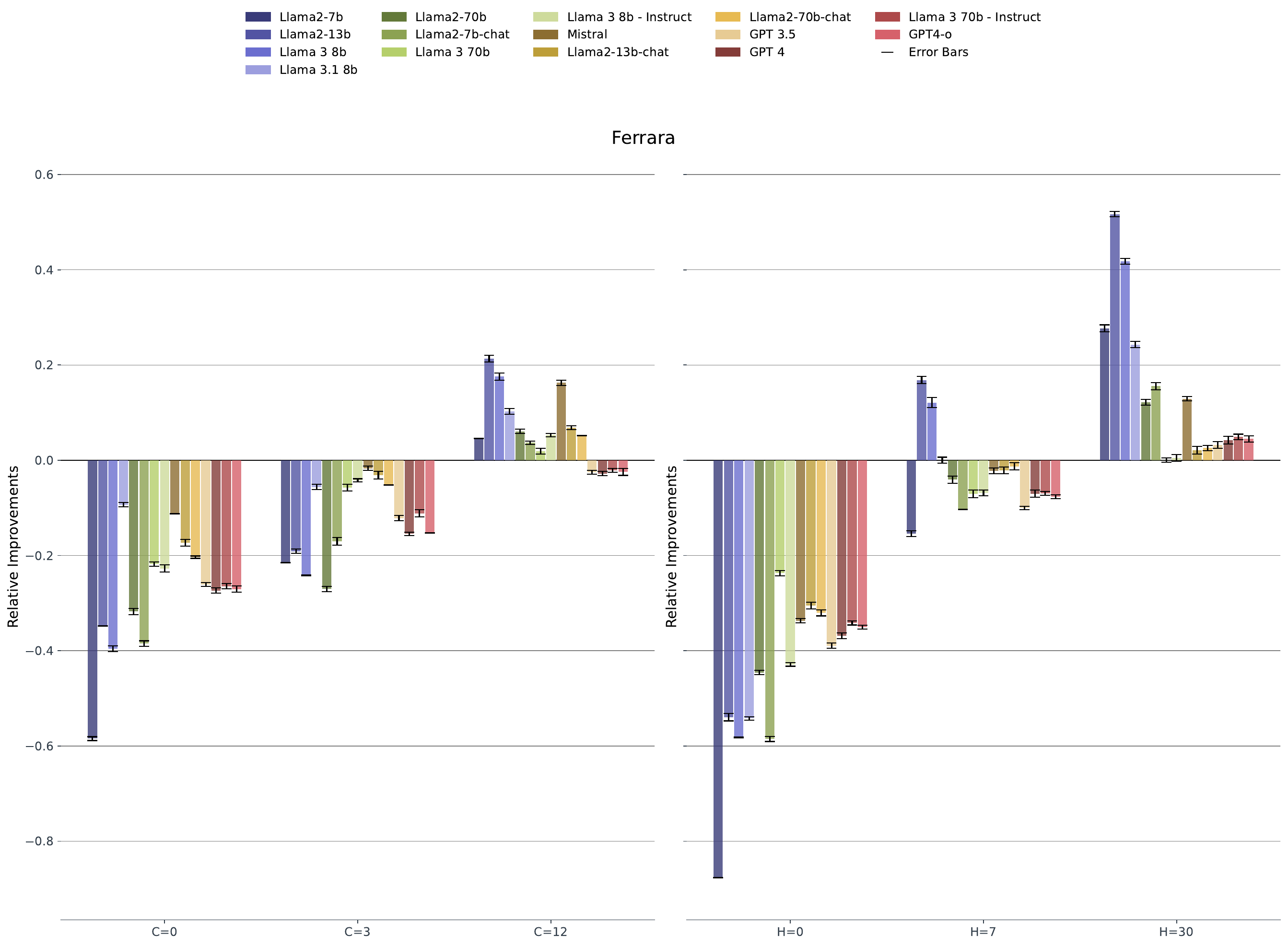}
    \caption{ Results in terms of relative improvements and drops for the scenario with $C=6, H=15$ for the city of Ferrara when we modify the availability of contextual $C$ (left) and historical $H$ (right) information.}
    \label{fig:conf_c_h}
\end{figure}

\newpage
\section{Quiz for Data Contamination}
\label{app:Quiz}

An example of the quiz to test data contamination inspired by \cite{golchin2023data}. To construct the questions, we extracted a random raw from the original dataset and assigned a random letter between A, B, C and D. All the other options are slightly modified to provide the model with realistic but non-existing entries of the dataset. For example, we change the user identifier (first value), the category of the location or, the last few characters of the location identifier. The LLMs were supposed to select the correct option given their potential knowledge of the dataset. It turned out that any of the LLMs were consistently selecting the right options. For instance, GPT-3.5 selected the right option 9 times over 50 quizzes and represented the best-performing model. 

\begin{tabular}{|p{\textwidth}|}
\hline
\noindent\textbf{Instruction:} You are provided with a four-choice quiz. Your task is to correctly select the option corresponding to an instance from the Foursquare NYC (``dataset\_TSMC2014\_NYC.txt'') dataset. \\

\noindent When selecting the option, you must ensure that you follow the following rules: \\
\begin{enumerate}
    \item You must ensure that you only generate a single option letter as your answer.
    \item If you do not know the dataset or the correct answer, you must select option ``E) None of the provided options.''
\end{enumerate} \\

\noindent \textit{Hint: While all the following options seem similar, there is only one option that reflects an exact match with respect to the original instance.} \\

\textbf{Options}: \\
\begin{itemize}
    \item A) 390 \, 4bcde547511f95210d62b5c7 \, 4bf58dd8d48988d124941735 \, Office \, 40.750945522488436 \, -74.00563392176072 \, -240 \, Tue Apr 03 18:15:07 +0000 2012
    \item B) 390 \, 44af9feef964a5202b351fe3 \, 4bf58dd8d48988d1c1941735 \, Mexican Restaurant \, 40.747738169430534 \, -73.98519814526952 \, -192 \, Tue Apr 03 18:15:33 +0000 2012
    \item C) 390 \, 44af9feef964a5202b351fe3 \, 4bf58dd8d48988d1c1910101 \, Mexican Restaurant \, 40.747738169430534 \, -73.98519814526952 \, -192 \, Tue Apr 03 18:15:33 +0000 2012
    \item D) 390 \, 44af9feef964a5202b351fe3 \, 4bf58cc8d48988d1c1941735 \, Office \, 40.747738169430534 \, -73.98519814526952 \, -192 \, Tue Apr 03 18:15:33 +0000 2012
    \item E) None of the provided options.
\end{itemize} \\
\hline
\end{tabular}

\end{document}